\documentclass[aps,twocolumn,showpacs,superscriptaddress,10pt]{revtex4-1}
\usepackage{graphicx}
\usepackage{color}

\begin{document}

\title{Tuning Kerr-Soliton Frequency Combs to Atomic Resonances}
% Microresonator Kerr Soliton Frequency Comb for Atomic Spectroscopy
% Tuning Kerr-soliton frequency combs for atomic spectroscopy
\author{Su-Peng Yu}
\email{supeng.yu@nist.gov}
\affiliation{Time and Frequency Division, NIST, Boulder, Colorado, USA}
\affiliation{Department of Physics, University of Colorado, Boulder, Colorado, USA}
\author{Travis C. Briles}
\author{Gregory T. Moille}
\author{Xiyuan Lu}
\affiliation{Center for Nanoscale Science and Technology, NIST, Gaithersburg, Maryland, USA}
\author{Scott A. Diddams}
\affiliation{Time and Frequency Division, NIST, Boulder, Colorado, USA}
\affiliation{Department of Physics, University of Colorado, Boulder, Colorado, USA}
\author{Kartik Srinivasan}
\affiliation{Center for Nanoscale Science and Technology, NIST, Gaithersburg, Maryland, USA}
\author{Scott B. Papp}
\email{scott.papp@nist.gov}
\affiliation{Time and Frequency Division, NIST, Boulder, Colorado, USA}
\affiliation{Department of Physics, University of Colorado, Boulder, Colorado, USA}

%\author{Su-Peng Yu}
%\author[1,2]{Travis C. Briles}
%\author[3]{Gregory T. Moille}
%\author[3]{Xiyuan Lu}
%\author[1,2]{Scott A. Diddams}
%\author[3]{Kartik Srinivasan}
%\author[1,2,*]{Scott B. Papp}

%\affil[1]{Time and Frequency Division, NIST, Boulder, Colorado, USA}
%\affil[2]{Department of Physics, University of Colorado, Boulder, Colorado, USA}
%\affil[3]{Center for Nanoscale Science and Technology, NIST, Gaithersburg, Maryland, USA}
%\affil[*]{Correspondence to: email@my-email.com and scott.papp@nist.gov}

% To be edited by editor
% \dates{Compiled \today}

%\ociscodes{(120.4800) Optical standards and testing; (130.0130) Integrated optics; (190.0190) Nonlinear optics; (190.5530) Pulse propagation and temporal solitons.}

%120.4800, 130.0130, 190.0190, 190.5530
% To be edited by editor % \doi{\url{http://dx.doi.org/10.1364/optica.XX.XXXXXX}}

\begin{abstract}
Frequency combs based on nonlinear-optical phenomena in integrated photonics are a versatile light source that can explore new applications, including frequency metrology, optical communications, and sensing. We demonstrate robust frequency-control strategies for near-infrared, octave-bandwidth soliton frequency combs, created with nanofabricated silicon-nitride ring resonators. Group-velocity-dispersion engineering allows operation with a 1064 nm pump laser and generation of dual-dispersive-wave frequency combs linking wavelengths between approximately 767 nm and 1556 nm.  To tune the mode frequencies of the comb, which are spaced by 1 THz, we design a photonic chip containing 75 ring resonators with systematically varying dimensions and we use 50 $^\circ$C of thermo-optic tuning. This single-chip frequency comb source provides access to every wavelength including those critical for near-infrared atomic spectroscopy of rubidium, potassium, and cesium.  To make this possible, solitons are generated consistently from device-to-device across a single chip, using rapid pump frequency sweeps that are provided by an optical modulator. 
\end{abstract}

%\setboolean{displaycopyright}{False}

\maketitle
\section{Introduction}
Recently, there has been significant effort to develop chip-scale and integrated-photonics light sources and devices through nonlinear optics \cite{moss2013new}. These have ranged from the development of ultra-stable CW lasers based on stimulated Brillouin scattering \cite{li2013microwave,loh2015dual} to metrology with frequency combs based on supercontinuum generation in photonic waveguides \cite{carlson2017self,lamb2018optical} to dissipative-Kerr-soliton (DKS) formation in microresonators \cite{kippenberg2018dissipative}. Such soliton microcombs exhibit interesting nonlinear optical phenomena, including breathing oscillations \cite{bao2016observation,lucas2017breathing, yu2017breather}, dark-pulse formation \cite{xue2015mode} and soliton crystallization \cite{cole2017soliton}, and have enabled diverse applications like %they have found
%dual-comb spectroscopy \cite{suh2016microresonator,dutt2018chip}, 
communications \cite{marin2017microresonator}, %distance measurements \cite{suh2018soliton,trocha2018ultrafast}, 
%astronomical spectrograph calibration \cite{obrzud2017microphotonic,suh2018searching}, 
and optical-frequency synthesis \cite{spencer2018integrated}.  DKS combs have been demonstrated with a variety of materials, including fused silica fiber and microresonators \cite{yi2015soliton,wang2017universal}, crystalline magnesium fluoride \cite{matsko2016optical} and silicon nitride (Si$_3$N$_4$ hereafter SiN)\cite{brasch2016photonic, jaramillo2015deterministic, briles2018interlocking}. SiN is an appealing material for broad bandwidth combs due to its moderately high Kerr nonlinearity and low optical propagation loss. Additionally, it is compatible with standard lithographic fabrication techniques, allowing for reliable fabrication of resonators and integration with other photonic elements \cite{moss2013new}, providing a path for broadband, low-power consumption frequency combs on a photonic chip \cite{stern2018battery}.

Realizing DKS frequency combs that interface with optical-atomic transitions opens up timekeeping and other measurement applications. However alkali or alkaline-earth atomic resonances are largely prevalent in the 700 to 1000 nm spectral region, whereas many DKS experiments have operated in the 1550 nm band due to the availability of pump lasers, optical amplifiers, filters, detectors, and other components. Previous optical-clock demonstrations with soliton microcombs were pumped at 1550 nm \cite{papp2014microresonator}, and required second-harmonic generation to access atomic transitions. Moreover, atomic transitions are spectrally narrow and thus require finely tuning a frequency comb to reach resonance.

A central facet of designing DKS combs for the near-infrared \cite{Karpov2018} is that the very high microwave-to-THz repetition frequencies of microcombs and their ultrabroad-bandwidth spectra are almost exclusively determined by the geometry of the devices. A recipe has emerged to design and realize soliton microcombs at 1550 nm for some applications like optical-frequency synthesis \cite{briles2018interlocking}. The first step in designing the soliton spectrum is to select device geometries with anomalous group-velocity-dispersion (GVD) near the chosen pump wavelength \cite{agha2009theoretical}. A second step is to design higher than 2$^{\textnormal{nd}}$- order GVD contributions for generation of dispersive waves (DWs) \cite{okawachi2014bandwidth}, which lead to a localized increase in optical power at the edges of the soliton spectrum. A third step is tuning the carrier-envelope offset frequency of the soliton microcomb for compatibility with electronic detection on available highspeed photodetectors with bandwidth $\leq$ 20GHz.

\begin{figure}[!t]
\centering\includegraphics[width=8cm]{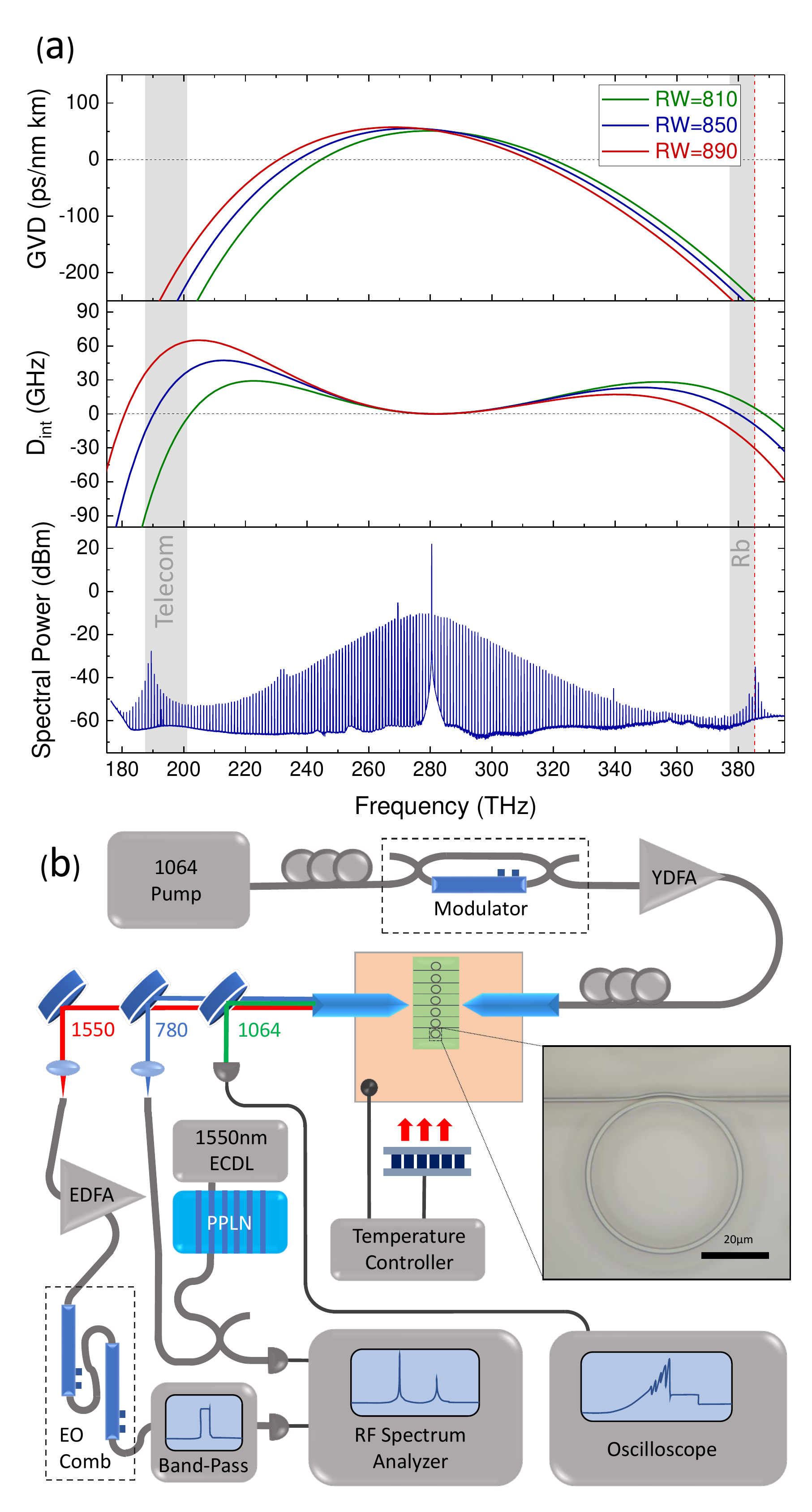}
\caption{Resonator dispersion design and experimental setup.  (a) The calculated GVD profile (top) and integrated modal dispersion (middle) for ring resonators with material thickness 667nm, radius of 22.5 $\mu$m and RW of 810 nm, 850 nm, and 890 nm.  The experimentally generated single soliton spectrum from the RW=850 nm device (bottom) has its strongest SDW line aligned to the Rb 2-h$\nu$ transition (red dashed line). (b) The experimental setup for soliton generation and comb diagnostics. An ECDL is coupled to a Mach-Zehnder intensity-modulator, amplified in a YDFA and coupled into the photonic chip using lensed fibers and precision positioning stages. The generated comb light is separated into spectral bands centered at 780, 1064 and 1550 nm for subsequent diagnostics. Each branch can be monitored in real time or undergo heterodyne with local oscillators. The notch filter used in the 1064 nm branch to reject pump light when monitoring soliton generation, is not shown.  The long-wavelength DW branch is amplified using EDFA to provide optical power for further measurement steps.}
\label{Fig:Fig1}
\end{figure}

Here, we explore frequency control and tuning of octave-spanning, near-infrared (NIR) soliton microcombs pumped at 1064 nm, towards use in atomic spectroscopy. We demonstrate a combination of GVD control of the comb's spectrum and thermo-optic frequency tuning for a single-chip solution to interface with rubidium (780 nm), cesium (852 nm), and potassium (767 nm) atoms. The advantage of a 1064 nm pump laser is that it bisects the frequency span between telecom wavelengths (1300 to 1600 nm) and atomic species that are commonly used for atomic clocks like strontium (699 nm) and calcium ion (729 nm) \cite{diddams2004standards, ludlow2015optical, Targat2013}. Our soliton microcomb provides up to 0.3$\mu$W per mode over the 760 to 900 nm range of interest for atomic spectroscopy, and also few $\mu$W per mode in the Telecom bands. The comb is pumped with standard 1064 nm laser technology, and we demonstrate high coherence with the soliton microcomb through detection of its 1 THz repetition frequency and $f$-$2f$ measurements.  

\section{Soliton Microcomb Design}
 
\begin{figure*}[!t]
\centering\includegraphics[width=16.5cm]{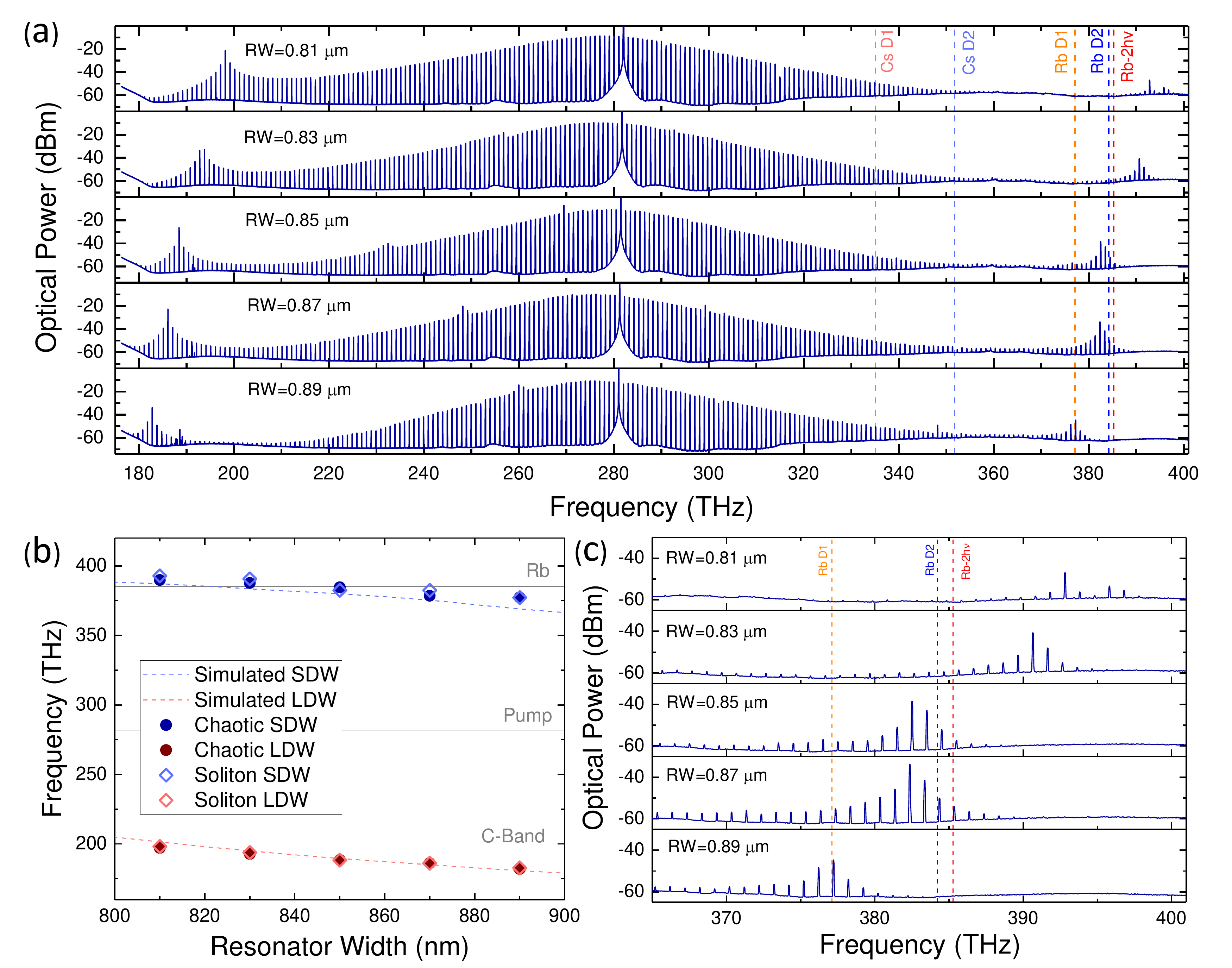}
 \caption{Soliton microcomb spectrum control through nanofabrication. (a) Single-soliton spectra obtained from resonators with varying waveguide widths on the same chip.  Relevant atomic transitions are indicated on the plot. (b) Comparison between measured dispersive wave frequencies and that predicted by simulated waveguide dispersion show good agreement within fabrication imprecision of resonator dimensions. (c) A zoom in plot near the rubidium optical transition frequencies.}
\label{Fig:Fig2}
\end{figure*}

The photonic-chip resonators in this study are SiN microring resonators embedded in a silicon-dioxide cladding. A ring radius of 22.5 $\mu$m (center of waveguide) is chosen to give a free-spectral range (FSR) of $\approx$ 1 THz, which reduces the comb operating power \cite{briles2018interlocking}. The devices are fabricated at wafer scale by Ligentec, using LPCVD stoichiometric silicon-nitride deposition, DUV stepper lithography, dry chemical etching, and separation to individual chips. Broadband, weakly anomalous GVD is created near the 1064 nm pump laser wavelength by design of the resonator dimensions, namely the SiN layer thickness and resonator waveguide width (RW). GVD simulations for resonators with SiN layer thicknesses of 667 nm and RW = 810, 850 and 890 nm illustrate the soliton microcomb design process; see the top panel of Fig. \ref{Fig:Fig1}(a). We indicate the telecom bands (1500-1600 nm) and the NIR range relevant for rubidium atomic spectroscopy (778-795 nm) with vertical gray boxes.

Our resonators also support the generation of DWs that extend the soliton spectrum outside the anomalous-GVD region.  The emitted DWs correspond approximately to the wavelengths of zero integrated dispersion, $D_{int}$ \cite{brasch2016photonic}.  The integrated dispersion is equivalent to the detuning of the cold-resonator frequencies $\nu_{\mu}$ from an equidistant grid centered about the pump laser and spaced by the free spectral range (FSR) about the pump laser,
\begin{equation}
D_{int} (\mu)=\nu_{\mu} - \left(\nu_p - \mu \times \frac{D_1}{2\pi}\right)
\end{equation}
\noindent Here, $\nu_p$ is the pump laser frequency, $\frac{D_1}{2\pi}$ is the resonator FSR near the pump and $\mu$ is an integer indexing the resonances relative to the pumped mode $\mu=0$. This quantity, given in units of GHz, is plotted in the middle panel of Fig. \ref{Fig:Fig1}(a) for the three corresponding traces in the top panel. We note that increasing the resonator waveguide-width (RW) parameter leads to an overall redshift of the GVD profile as well as the wavelengths of $D_{int}=0$. Since both the device layer thickness and RW affect the GVD profile, we can quasi-independently design and control the peak wavelength of the two DWs.

The linear approximation for the DW wavelengths given by $D_{int} (\mu) = 0$ neglects higher-order effects arising from soliton self-phase modulation \cite{skryabin2017self}.  However, it is a reasonable criteria for predicting the performance of resonator designs given the fabrication tolerances of $\sim$5 nm in device layer thickness and $\sim$20 nm in pattern dimensions, including effects of waveguide sidewall angle. These small deviations are accounted for by fabricating resonators of various RW values, stepped by 10's of nanometers, on a single chip. We design chips composed of $\sim$75 individual ring resonators to systematically cover the microcomb design space.

The bottom panel of Fig. \ref{Fig:Fig1}(a) shows the result of controllably taking advantage of the microcomb design space in SiN. The soliton microcomb spectrum that we present features a long-wavelength DW in the telecom band and short-wavelength DW with its strongest line aligned to the Rb 2-photon clock transition at 778 nm \cite{hilico1998metrological, zameroski2014pressure, bernard2000absolute}.% and a long-wavelength DW covering the telecom C band.

\section{Optical Measurement}

The setup we use for our experiments is shown in Fig. \ref{Fig:Fig1}(b). In particular, the apparatus is composed of elements for soliton microcomb generation and thermo-optic tuning, and for coherence measurements of the soliton microcomb's carrier-envelope-offset frequency and repetition frequency. The pump laser we use is an external cavity diode laser, providing coarse tuning from 1020 to 1070 nm to address a few of the TE1 ring resonator modes that are spaced by 1 THz.

In order to provide laser light to our resonators, coupling waveguides are fabricated in the silicon nitride device layer. Inverse tapering of the coupling waveguide at chip edge enables approximately 3 dB per facet insertion loss from standard lensed fibers. A `pulley' design \cite{hosseini2010systematic, li2015octave,briles2018interlocking} for the coupling waveguide achieves adequate coupling rates for low-power operation of the comb and also at the edges of the spectrum for output-coupling of the dispersive waves. We include a variation in the coupling gap between the pulley coupler and the resonator to achieve critical coupling, namely, to set the coupling strength between the resonator and the coupling waveguide to match the internal loss of the resonator. This coupling condition approximately maximizes the intra-resonator power for a given pump power, reducing the pump power required to support soliton microcombs. We extract information on the intrinsic and coupled quality factor by measuring the resonance linewidth and minimum transmission as a function of gap. The spectra shown in Figs \ref{Fig:Fig1}(a) and \ref{Fig:Fig2} were obtained using resonators with intrinsic quality factors of $\approx 1\times10^6$ and comb generation threshold powers as low as 9 mW in the coupling waveguide. Ongoing experiments show that intrinsic Q's up to $\approx 3\times10^6$ and threshold powers of 1 mW are possible at 1064 nm.

We efficiently generate soliton microcombs by controlling the pump-resonator detuning at a rate faster than the resonator thermal dynamics \cite{briles2018thermal,stone2018thermal,li2017stably}. This is implemented using an ultrafast electro-optic frequency shifting technique similar to that reported in Ref \cite{briles2018interlocking}. Here we utilize a standard lithium-niobate waveguide Mach-Zehnder intensity modulator in place of the single-sideband, suppressed-carrier modulator. The output of the intensity modulator is passed to a polarization controller and amplified in an ytterbium-doped fiber amplifier (YDFA) before being coupled into the chip.  The blue-detuned sideband frequency is tuned to resonance and then decreased by $\approx 5$ GHz in 100 ns.  This rapid pump frequency tuning technique minimizes the thermo-optic shift of the resonance frequency, which can destabilize the soliton. The red sideband and residual carrier are far detuned (by $\geq$20 resonator linewidths) from any resonances and do not adversely affect soliton generation. We filter out the pump light to monitor the generated comb power. The emergence of soliton microcombs from chaotic states as we vary the detuning is identified from the photodetected comb power, using an oscilloscope; see the trace plotted in Fig. \ref{Fig:Fig1}(b). 

\begin{figure}[!h]
\centering\includegraphics[width=9cm]{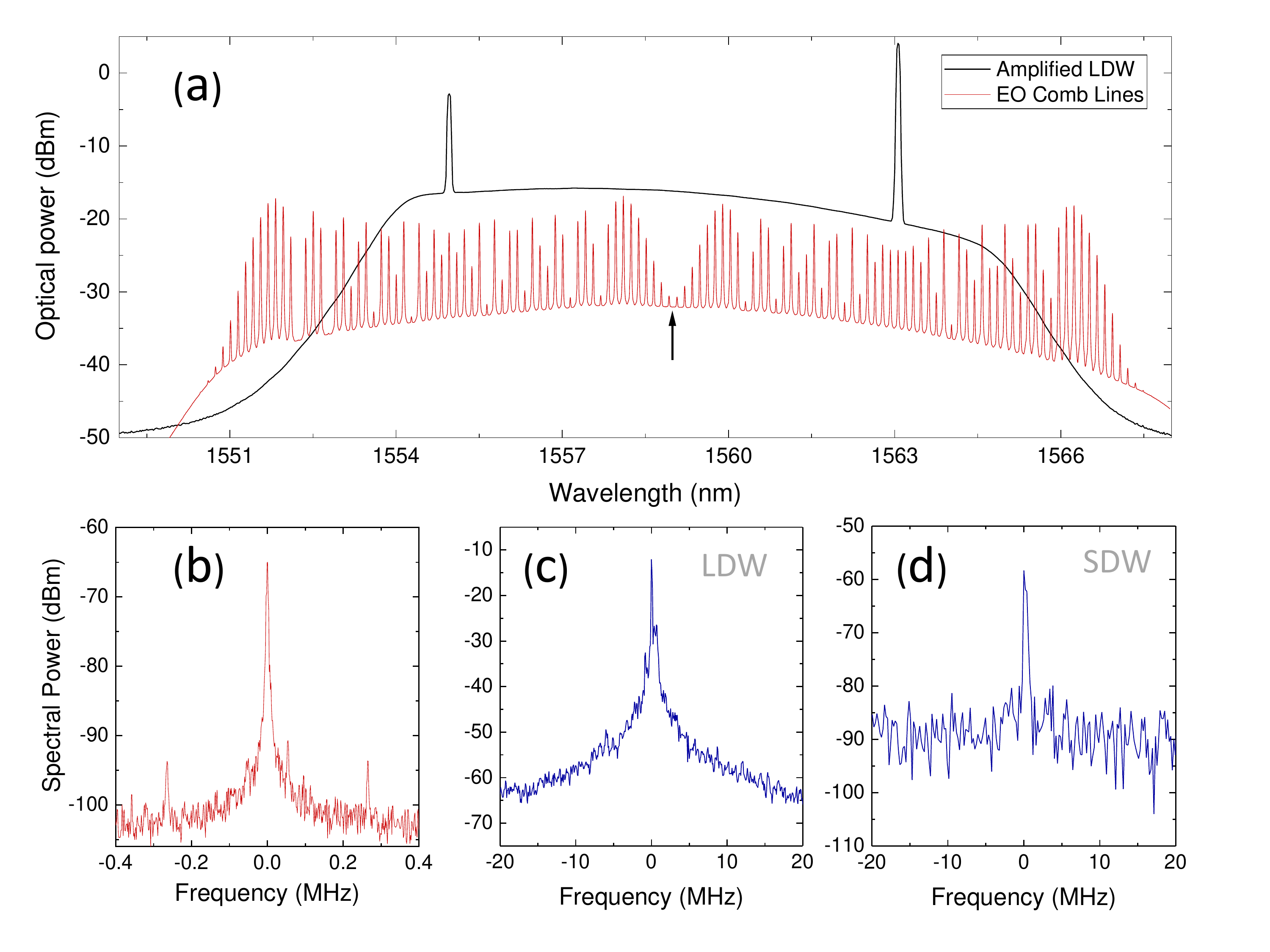}
\caption{Soliton coherence measurements.  (a) The filtered and amplified long-wavelength DW before (black) and after (red) being sent to electro-optic phase modulators.  A heterodyne beat is made at the wavelength marked with an arrow to evaluate fluctuations in THz comb mode spacing. This beat is shown in (b).  (c,d) Heterodyne beats of each DW with a low-noise auxiliary laser: (c) LDW at 1563 nm and (d) SDW at 777.3 nm. The traces (b) through (d) are taken with resolution bandwidth 1 kHz, 100 kHz, and 300 kHz, respectively.}
\label{Fig:Fig3}
\end{figure}

Optical spectra of single-soliton microcombs that we create are shown in Fig. \ref{Fig:Fig2}(a). We present spectra for five RW values, spaced by 20 nm from RW = 810-890 nm.  Each spectrum features two prominent dispersive waves on either side of the pump laser.  The measured DW frequencies in the telecom C/L band and the NIR 780 nm band are plotted versus RW in Fig. \ref{Fig:Fig2}(b) as hollow red and blue diamonds, respectively. For comparison, we also present DW frequency measurements obtained with blue-detuned chaotic combs (solid circles). This thorough and unique comparison of microcomb spectra is enabled by precision fabrication of numerous devices and reliable soltion microcomb generation.  Our measurements of the DW frequencies agree with our designs, which are based on finite-element calculations of $D_{int}$ (open circles), using accurate Sellmeier coefficients for SiN \cite{Luke2015} and silica \cite{Malitson1965} and a fabrication precision that we assess to be $\sim$20 nm. The predictability of dispersive wave frequencies enables placement of dispersive waves on the targeted Rb resonances, as shown in the zoom-in plot in Fig. \ref{Fig:Fig3}(c).

We measure the repetition frequency of the soliton microcomb by electro-optic modulation of the LDW near 1550 nm.  Such measurements are shown for the RW=0.83 $\mu$m device in Fig. \ref{Fig:Fig3}. This procedure creates a low frequency $<$100 MHz optical heterodyne beatnote $f_{beat}$ that represents the 0.999291 THz repetition frequency. To facilitate these measurements, we use optical filters to isolate the dispersive waves near 1550 nm and 780 nm. The long-wavelength DW is amplified with a C-band erbium-doped fiber amplifier (EDFA), filtered, and sent to two electro-optic phase modulators operating in series and driven at $f_{eo}\approx 16.65$ GHz to create many coherent sidebands around each comb tooth \cite{del2012hybrid}. We detect $f_{beat}$ by photodetecting the phase modulation sidebands nearly in between the two soliton microcomb modes, allowing the repetition frequency to be determined via $f_{rep}=60\cdot f_{eo}+f_{beat}$. A narrow optical bandpass filter selects the desired interfering modes (marked with an arrow in Fig. \ref{Fig:Fig3}(a)) before photodetecting $f_{beat}$.  The highly coherent beat obtained from this measurement (shown in Fig. \ref{Fig:Fig3}(b)) confirms low-noise soliton operation.  
 
We evaluate the frequency noise on individual comb modes by recording optical heterodyne beats between the soliton comb and an auxiliary low-noise CW laser.  We focus on large comb mode numbers at a large frequency offset from the 1064 nm pump laser, which are interesting to test since they are most sensitive to multiplicative noise of the comb. A C-band external cavity diode laser is heterodyned directly with the long-wavelength DW near 1563 nm, and it is also frequency doubled in a bulk periodically poled lithium niobate crystal for the short-wavelength DW near 777.3 nm. For both heterodyne beatnotes, which are shown in Figs. \ref{Fig:Fig3} (c) and (d) for the long- and short-wavelength DW, respectively, we observe linewidths of a few megahertz that we associate with the linewidth of the soliton microcomb modes. This is reasonable considering that the free-running repetition rate has a frequency linewidth of $\approx$5 kHz and the comb modes we characterize in Figs. \ref{Fig:Fig3} (c) and (d) are $\approx$100 modes from the pump.

We note that the DW wavelengths were chosen in this case to be near-harmonic to facilitate $f$-$2f$ referencing. For this particular device, the offset frequency $f_0$ is $\approx 80 \pm 1$ GHz. While such a high frequency makes a direct electronic measurement of $f_0$ challenging, in future designs we would systematically vary the resonator radius in steps of $\approx 10$ nm to create a specific device that satisfies the conditions $f_0 \leq 20$ GHz \cite{briles2018interlocking}.  

%and a heterodyne is recorded and then frequency doubled for the short-wavelength DW) is tuned near the DW to generate RF beatnotes, which show few-MHz linewidth, likely contributed from fluctuation in repetition rate. \textcolor{red}{work on this!}.

%noise on  noise We further measure the properties of the two DWs by heterodyne measurements.%The sidebands introduced by the modulator spans across the 1 THz FSR of the resonators. A narrow band-pass filter selects the modes where the two sets of sidebands overlap for photodetection. 
%The EO Comb produces equi-distance sidebands defined by integer multiples of the RF driving frequency. By measuring the beatnote between sidebands produced from two adjacent modes of the resonator, the 1 THz repetition rate of the micro-resonators can be measured efficiently.

\section{Thermo-optic Fine Tuning}
%Fig. 4a, 4c are from March 13 (I think) And this is nominal RW=0.89um (or actual RW of 0.87um which is the value quoted in the paper)%Fig. 4b comes april 13 data.  

\begin{figure}[h!]
\centering\includegraphics[width=9cm]{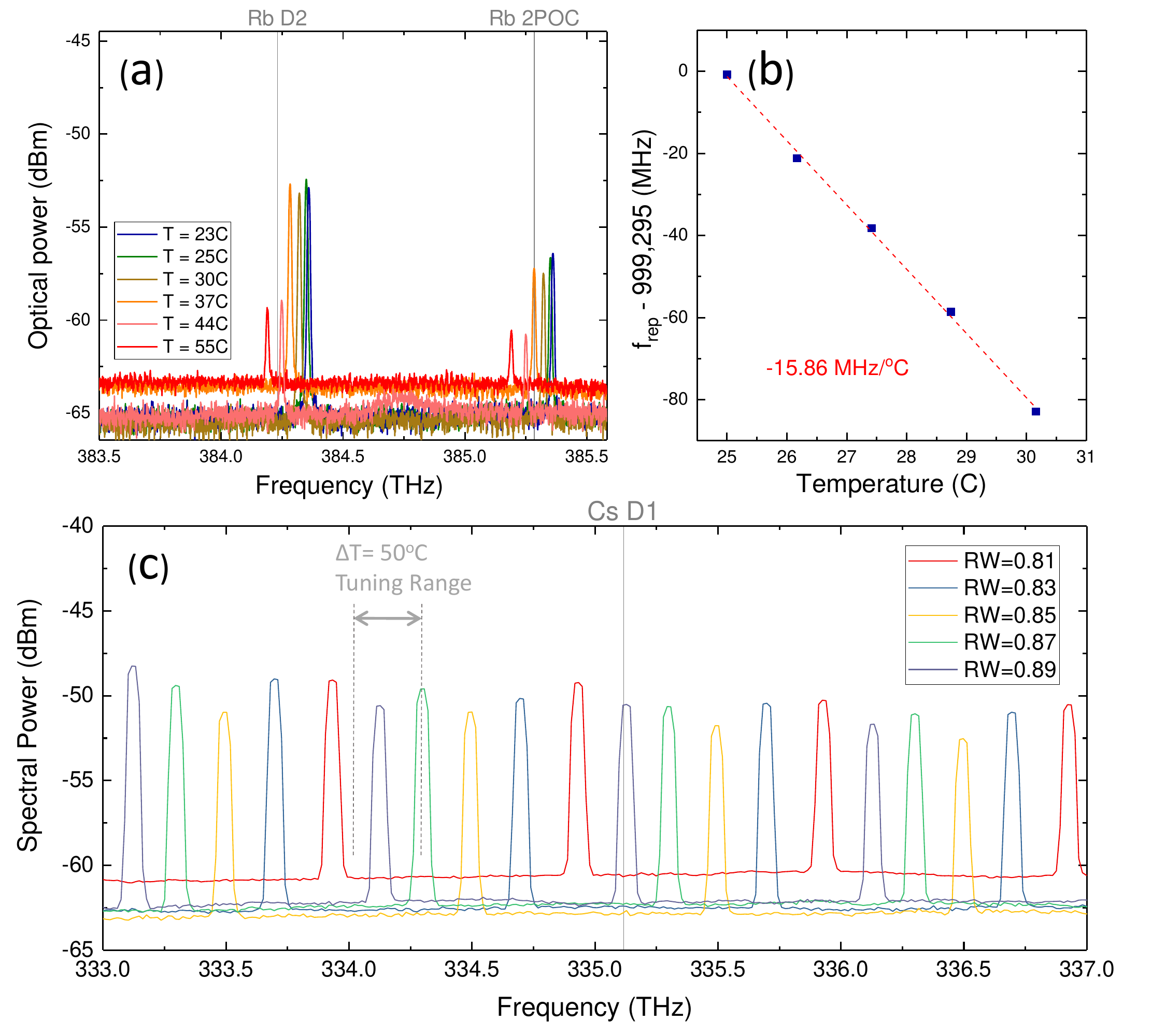}
\caption{Thermo-optic tuning of soliton spectrum.  (a) The thermal tuning of two modes of the soliton spectrum across the Rb two-photon and Rb D2 transitions. (b) The repetition rate of the soliton can also be thermally tuned. (c) Overlay of the spectra presented in Fig. \ref{Fig:Fig2} plotted over the range from 333 to 337 THz.  Note that any arbitrary frequency can be reached with the current RW sweep and a 50 $^\circ$C thermal tuning range.}
\label{Fig:Fig4}
\end{figure}

We finely shift the absolute frequencies of the soliton comb modes beyond the level provided by nanofabrication process control, using thermo-optic tuning of the resonator.  This approach combines the high optical power per mode of DKS combs with the high spectral resolution typical of lower repetition rate combs \cite{Heinecke2009}. The temperature of the chip is set with a PID-controlled thermoelectric heater; see Fig. \ref{Fig:Fig1} (b) for the setup.  Fig \ref{Fig:Fig4} (a) and (b) analyze the thermal tuning of solitons created in RW = 0.87 $\mu$m resonators (second to last panel in Fig. \ref{Fig:Fig2} (a) and (c)).  Comb modes near 778 nm have a measured tuning coefficient of -5.48 GHz/ $^\circ$C which is sufficiently large to access the Rb $2h\nu$ and D2 resonances at chip temperatures of $\approx 37$ and $\approx 44$ $^\circ$C respectively; see Fig. \ref{Fig:Fig4}(a).  The change in the comb modes power between these two temperatures is associated with hysteresis in switching behavior between distinct operating regimes of the single soliton pulse, which we plan to explore in more detail in future work. The measured linear thermal tuning of the soliton repetition rate of $15.86\pm 0.34$ MHz $/^\circ C$ shown in Fig. \ref{Fig:Fig4} (b) amounts to a fractional change of $\approx 16$ ppm on the $1.0$ THz spacing. This is in reasonable agreement with measured thermo-optic coefficients of bulk silicon nitride at 20 to 30 ppm \cite{ikeda2008thermal}%\textcolor{red}{double check this reference! ...I looked, it seems they quoted 10ppm for PECVD SiN?} 
as well as experimental measurements on resonators pumped at 1550 nm \cite{xue2016thermal}. In future work we plan to incorporate integrated heaters \cite{xue2016thermal,joshi2016thermally}, which will reduce the overall electrical power consumption.
 
 %In Fig. \ref{Fig:Fig4}(c) we overlay the spectra from Fig. \ref{Fig:Fig2}. Given the thermal tuning rate of -5.48 GHz/ $^\circ$C, we can span the frequency differences between two consecutive devices on one chip within a 50  $^\circ$C temperature range.
%We found such number of available device allows for reaching of arbitrary desired wavelength in the design bandwidth with an individual chip. A combination of fine device geometric parameter sweep with precise thermo-optic control for fine-tuning is sufficient to generate a comb line at any desired wavelength.

Through both lithographic variation in resonator dimensions and thermo-optic tuning, we exert comprehensive frequency control over the SiN comb. A typical chip used in this study contains 70 devices, which is sufficient to include combinations of fine resonator width and gap sweeps for desired dispersion and coupling. A modest $\Delta$T = $50^{\circ} C$ thermal-optic tuning of the chip is sufficient to place a comb line onto any desired wavelength within the design bandwidth of the resonators. For example, to probe the cesium D1 transition, we can select a device with ring width 870 nm or 890 nm, and thermally tune the modes toward lower frequency to align to the desired transition, as shown in Fig. \ref{Fig:Fig4}(c).
 
\section{Conclusion}
We have demonstrated optical-frequency-comb generation in the near infrared, using silicon nitride ring resonators. These devices demonstrate octave-span comb generation that can address optical resonances of various atomic species. Sufficient power output in the far short- and long-wavelength sides of the comb spectra is achieved utilizing engineered dispersive waves and efficient out-coupling waveguides. Single Kerr solitons are reliably generated, using a fast frequency sweep technique that works with all devices presented in this study, and the coherent properties of the soliton microcombs are verified by heterodyne beatnote measurements.

From the perspective of building a capable frequency-comb system, our 1064 nm pumped soliton microcombs could be combined with mature optical amplifier, optical modulator, and integrated second-harmonic generation technology in the 1550 nm band. These tools present a route to $f$-$2f$ detection of our combs that are frequency tunable continuously across the near infrared. Indeed, we have demonstrated repetition rate detection of the soliton microcomb with phase modulation. Further, we have demonstrated the flexibility of our system by thermally tuning optical comb modes onto Rb and Cs resonances, which would be enabling for future experiments that explore spectroscopy and metrology with atomic resonances.

%On the long-wavelength side of the generated spectra, we again benefit from mature optical amplifier (erbium ion, 1500 to 1600 nm) and second-harmonic generation technologies that will facilitate self-referencing \cite{briles2018interlocking}. The amplified long-wavelength DW is readily available for frequency-doubling for $f$-$2f$ self-reference techniques. 

\section*{Funding Information}
Funding provided by DARPA ACES and DODOS, Air Force Office of Scientific Research (FA9550-16-1-0016), NIST, NASA, and NRC.

\section*{Acknowledgments}

%Formal funding declarations should not be included in the acknowledgments but in a Funding Information section as shown above. The acknowledgments may contain information that is not related to funding:

We thank Jordan R. Stone and Connor Fredrick for a careful reading of the manuscript and Ligentec for fabrication of the optical resonators. This work is a contribution of the U.S. Government and is not subject to copyright. Mention of specific companies or trade names is for scientific communication only, and does not constitute an endorsement by NIST. %(Do we need to add the CNST toolbox authors here?)

%\bigskip \noindent See \href{link}{Supplement 1} for supporting content.

% Bibliography
\bibliography{brilesBIBv7,1064CombRef}

%merlin.mbs apsrev4-1.bst 2010-07-25 4.21a (PWD, AO, DPC) hacked
%Control: key (0)
%Control: author (8) initials jnrlst
%Control: editor formatted (1) identically to author
%Control: production of article title (-1) disabled
%Control: page (0) single
%Control: year (1) truncated
%Control: production of eprint (0) enabled
\begin{thebibliography}{43}%
\makeatletter
\providecommand \@ifxundefined [1]{%
 \@ifx{#1\undefined}
}%
\providecommand \@ifnum [1]{%
 \ifnum #1\expandafter \@firstoftwo
 \else \expandafter \@secondoftwo
 \fi
}%
\providecommand \@ifx [1]{%
 \ifx #1\expandafter \@firstoftwo
 \else \expandafter \@secondoftwo
 \fi
}%
\providecommand \natexlab [1]{#1}%
\providecommand \enquote  [1]{``#1''}%
\providecommand \bibnamefont  [1]{#1}%
\providecommand \bibfnamefont [1]{#1}%
\providecommand \citenamefont [1]{#1}%
\providecommand \href@noop [0]{\@secondoftwo}%
\providecommand \href [0]{\begingroup \@sanitize@url \@href}%
\providecommand \@href[1]{\@@startlink{#1}\@@href}%
\providecommand \@@href[1]{\endgroup#1\@@endlink}%
\providecommand \@sanitize@url [0]{\catcode `\\12\catcode `\$12\catcode
  `\&12\catcode `\#12\catcode `\^12\catcode `\_12\catcode `\%12\relax}%
\providecommand \@@startlink[1]{}%
\providecommand \@@endlink[0]{}%
\providecommand \url  [0]{\begingroup\@sanitize@url \@url }%
\providecommand \@url [1]{\endgroup\@href {#1}{\urlprefix }}%
\providecommand \urlprefix  [0]{URL }%
\providecommand \Eprint [0]{\href }%
\providecommand \doibase [0]{http://dx.doi.org/}%
\providecommand \selectlanguage [0]{\@gobble}%
\providecommand \bibinfo  [0]{\@secondoftwo}%
\providecommand \bibfield  [0]{\@secondoftwo}%
\providecommand \translation [1]{[#1]}%
\providecommand \BibitemOpen [0]{}%
\providecommand \bibitemStop [0]{}%
\providecommand \bibitemNoStop [0]{.\EOS\space}%
\providecommand \EOS [0]{\spacefactor3000\relax}%
\providecommand \BibitemShut  [1]{\csname bibitem#1\endcsname}%
\let\auto@bib@innerbib\@empty
%</preamble>
\bibitem [{\citenamefont {Moss}\ \emph {et~al.}(2013)\citenamefont {Moss},
  \citenamefont {Morandotti}, \citenamefont {Gaeta},\ and\ \citenamefont
  {Lipson}}]{moss2013new}%
  \BibitemOpen
  \bibfield  {author} {\bibinfo {author} {\bibfnamefont {D.~J.}\ \bibnamefont
  {Moss}}, \bibinfo {author} {\bibfnamefont {R.}~\bibnamefont {Morandotti}},
  \bibinfo {author} {\bibfnamefont {A.~L.}\ \bibnamefont {Gaeta}}, \ and\
  \bibinfo {author} {\bibfnamefont {M.}~\bibnamefont {Lipson}},\ }\href@noop {}
  {\bibfield  {journal} {\bibinfo  {journal} {Nature Photonics}\ }\textbf
  {\bibinfo {volume} {7}},\ \bibinfo {pages} {597} (\bibinfo {year}
  {2013})}\BibitemShut {NoStop}%
\bibitem [{\citenamefont {Li}\ \emph {et~al.}(2013)\citenamefont {Li},
  \citenamefont {Lee},\ and\ \citenamefont {Vahala}}]{li2013microwave}%
  \BibitemOpen
  \bibfield  {author} {\bibinfo {author} {\bibfnamefont {J.}~\bibnamefont
  {Li}}, \bibinfo {author} {\bibfnamefont {H.}~\bibnamefont {Lee}}, \ and\
  \bibinfo {author} {\bibfnamefont {K.~J.}\ \bibnamefont {Vahala}},\
  }\href@noop {} {\bibfield  {journal} {\bibinfo  {journal} {Nature
  Communications}\ }\textbf {\bibinfo {volume} {4}},\ \bibinfo {pages} {2097}
  (\bibinfo {year} {2013})}\BibitemShut {NoStop}%
\bibitem [{\citenamefont {Loh}\ \emph {et~al.}(2015)\citenamefont {Loh},
  \citenamefont {Green}, \citenamefont {Baynes}, \citenamefont {Cole},
  \citenamefont {Quinlan}, \citenamefont {Lee}, \citenamefont {Vahala},
  \citenamefont {Papp},\ and\ \citenamefont {Diddams}}]{loh2015dual}%
  \BibitemOpen
  \bibfield  {author} {\bibinfo {author} {\bibfnamefont {W.}~\bibnamefont
  {Loh}}, \bibinfo {author} {\bibfnamefont {A.~A.}\ \bibnamefont {Green}},
  \bibinfo {author} {\bibfnamefont {F.~N.}\ \bibnamefont {Baynes}}, \bibinfo
  {author} {\bibfnamefont {D.~C.}\ \bibnamefont {Cole}}, \bibinfo {author}
  {\bibfnamefont {F.~J.}\ \bibnamefont {Quinlan}}, \bibinfo {author}
  {\bibfnamefont {H.}~\bibnamefont {Lee}}, \bibinfo {author} {\bibfnamefont
  {K.~J.}\ \bibnamefont {Vahala}}, \bibinfo {author} {\bibfnamefont {S.~B.}\
  \bibnamefont {Papp}}, \ and\ \bibinfo {author} {\bibfnamefont {S.~A.}\
  \bibnamefont {Diddams}},\ }\href@noop {} {\bibfield  {journal} {\bibinfo
  {journal} {Optica}\ }\textbf {\bibinfo {volume} {2}},\ \bibinfo {pages} {225}
  (\bibinfo {year} {2015})}\BibitemShut {NoStop}%
\bibitem [{\citenamefont {Carlson}\ \emph {et~al.}(2017)\citenamefont
  {Carlson}, \citenamefont {Hickstein}, \citenamefont {Lind}, \citenamefont
  {Droste}, \citenamefont {Westly}, \citenamefont {Nader}, \citenamefont
  {Coddington}, \citenamefont {Newbury}, \citenamefont {Srinivasan},
  \citenamefont {Diddams} \emph {et~al.}}]{carlson2017self}%
  \BibitemOpen
  \bibfield  {author} {\bibinfo {author} {\bibfnamefont {D.~R.}\ \bibnamefont
  {Carlson}}, \bibinfo {author} {\bibfnamefont {D.~D.}\ \bibnamefont
  {Hickstein}}, \bibinfo {author} {\bibfnamefont {A.}~\bibnamefont {Lind}},
  \bibinfo {author} {\bibfnamefont {S.}~\bibnamefont {Droste}}, \bibinfo
  {author} {\bibfnamefont {D.}~\bibnamefont {Westly}}, \bibinfo {author}
  {\bibfnamefont {N.}~\bibnamefont {Nader}}, \bibinfo {author} {\bibfnamefont
  {I.}~\bibnamefont {Coddington}}, \bibinfo {author} {\bibfnamefont {N.~R.}\
  \bibnamefont {Newbury}}, \bibinfo {author} {\bibfnamefont {K.}~\bibnamefont
  {Srinivasan}}, \bibinfo {author} {\bibfnamefont {S.~A.}\ \bibnamefont
  {Diddams}},  \emph {et~al.},\ }\href@noop {} {\bibfield  {journal} {\bibinfo
  {journal} {Optics letters}\ }\textbf {\bibinfo {volume} {42}},\ \bibinfo
  {pages} {2314} (\bibinfo {year} {2017})}\BibitemShut {NoStop}%
\bibitem [{\citenamefont {Lamb}\ \emph {et~al.}(2018)\citenamefont {Lamb},
  \citenamefont {Carlson}, \citenamefont {Hickstein}, \citenamefont {Stone},
  \citenamefont {Diddams},\ and\ \citenamefont {Papp}}]{lamb2018optical}%
  \BibitemOpen
  \bibfield  {author} {\bibinfo {author} {\bibfnamefont {E.~S.}\ \bibnamefont
  {Lamb}}, \bibinfo {author} {\bibfnamefont {D.~R.}\ \bibnamefont {Carlson}},
  \bibinfo {author} {\bibfnamefont {D.~D.}\ \bibnamefont {Hickstein}}, \bibinfo
  {author} {\bibfnamefont {J.~R.}\ \bibnamefont {Stone}}, \bibinfo {author}
  {\bibfnamefont {S.~A.}\ \bibnamefont {Diddams}}, \ and\ \bibinfo {author}
  {\bibfnamefont {S.~B.}\ \bibnamefont {Papp}},\ }\href@noop {} {\bibfield
  {journal} {\bibinfo  {journal} {Physical Review Applied}\ }\textbf {\bibinfo
  {volume} {9}},\ \bibinfo {pages} {024030} (\bibinfo {year}
  {2018})}\BibitemShut {NoStop}%
\bibitem [{\citenamefont {Kippenberg}\ \emph {et~al.}(2018)\citenamefont
  {Kippenberg}, \citenamefont {Gaeta}, \citenamefont {Lipson},\ and\
  \citenamefont {Gorodetsky}}]{kippenberg2018dissipative}%
  \BibitemOpen
  \bibfield  {author} {\bibinfo {author} {\bibfnamefont {T.~J.}\ \bibnamefont
  {Kippenberg}}, \bibinfo {author} {\bibfnamefont {A.~L.}\ \bibnamefont
  {Gaeta}}, \bibinfo {author} {\bibfnamefont {M.}~\bibnamefont {Lipson}}, \
  and\ \bibinfo {author} {\bibfnamefont {M.~L.}\ \bibnamefont {Gorodetsky}},\
  }\href@noop {} {\bibfield  {journal} {\bibinfo  {journal} {Science}\ }\textbf
  {\bibinfo {volume} {361}},\ \bibinfo {pages} {eaan8083} (\bibinfo {year}
  {2018})}\BibitemShut {NoStop}%
\bibitem [{\citenamefont {Bao}\ \emph {et~al.}(2016)\citenamefont {Bao},
  \citenamefont {Jaramillo-Villegas}, \citenamefont {Xuan}, \citenamefont
  {Leaird}, \citenamefont {Qi},\ and\ \citenamefont
  {Weiner}}]{bao2016observation}%
  \BibitemOpen
  \bibfield  {author} {\bibinfo {author} {\bibfnamefont {C.}~\bibnamefont
  {Bao}}, \bibinfo {author} {\bibfnamefont {J.~A.}\ \bibnamefont
  {Jaramillo-Villegas}}, \bibinfo {author} {\bibfnamefont {Y.}~\bibnamefont
  {Xuan}}, \bibinfo {author} {\bibfnamefont {D.~E.}\ \bibnamefont {Leaird}},
  \bibinfo {author} {\bibfnamefont {M.}~\bibnamefont {Qi}}, \ and\ \bibinfo
  {author} {\bibfnamefont {A.~M.}\ \bibnamefont {Weiner}},\ }\href@noop {}
  {\bibfield  {journal} {\bibinfo  {journal} {Physical Review Letters}\
  }\textbf {\bibinfo {volume} {117}},\ \bibinfo {pages} {163901} (\bibinfo
  {year} {2016})}\BibitemShut {NoStop}%
\bibitem [{\citenamefont {Lucas}\ \emph {et~al.}(2017)\citenamefont {Lucas},
  \citenamefont {Karpov}, \citenamefont {Guo}, \citenamefont {Gorodetsky},\
  and\ \citenamefont {Kippenberg}}]{lucas2017breathing}%
  \BibitemOpen
  \bibfield  {author} {\bibinfo {author} {\bibfnamefont {E.}~\bibnamefont
  {Lucas}}, \bibinfo {author} {\bibfnamefont {M.}~\bibnamefont {Karpov}},
  \bibinfo {author} {\bibfnamefont {H.}~\bibnamefont {Guo}}, \bibinfo {author}
  {\bibfnamefont {M.}~\bibnamefont {Gorodetsky}}, \ and\ \bibinfo {author}
  {\bibfnamefont {T.~J.}\ \bibnamefont {Kippenberg}},\ }\href@noop {}
  {\bibfield  {journal} {\bibinfo  {journal} {Nature Communications}\ }\textbf
  {\bibinfo {volume} {8}},\ \bibinfo {pages} {736} (\bibinfo {year}
  {2017})}\BibitemShut {NoStop}%
\bibitem [{\citenamefont {Yu}\ \emph {et~al.}(2017)\citenamefont {Yu},
  \citenamefont {Jang}, \citenamefont {Okawachi}, \citenamefont {Griffith},
  \citenamefont {Luke}, \citenamefont {Miller}, \citenamefont {Ji},
  \citenamefont {Lipson},\ and\ \citenamefont {Gaeta}}]{yu2017breather}%
  \BibitemOpen
  \bibfield  {author} {\bibinfo {author} {\bibfnamefont {M.}~\bibnamefont
  {Yu}}, \bibinfo {author} {\bibfnamefont {J.~K.}\ \bibnamefont {Jang}},
  \bibinfo {author} {\bibfnamefont {Y.}~\bibnamefont {Okawachi}}, \bibinfo
  {author} {\bibfnamefont {A.~G.}\ \bibnamefont {Griffith}}, \bibinfo {author}
  {\bibfnamefont {K.}~\bibnamefont {Luke}}, \bibinfo {author} {\bibfnamefont
  {S.~A.}\ \bibnamefont {Miller}}, \bibinfo {author} {\bibfnamefont
  {X.}~\bibnamefont {Ji}}, \bibinfo {author} {\bibfnamefont {M.}~\bibnamefont
  {Lipson}}, \ and\ \bibinfo {author} {\bibfnamefont {A.~L.}\ \bibnamefont
  {Gaeta}},\ }\href@noop {} {\bibfield  {journal} {\bibinfo  {journal} {Nature
  Communications}\ }\textbf {\bibinfo {volume} {8}},\ \bibinfo {pages} {14569}
  (\bibinfo {year} {2017})}\BibitemShut {NoStop}%
\bibitem [{\citenamefont {Xue}\ \emph {et~al.}(2015)\citenamefont {Xue},
  \citenamefont {Xuan}, \citenamefont {Liu}, \citenamefont {Wang},
  \citenamefont {Chen}, \citenamefont {Wang}, \citenamefont {Leaird},
  \citenamefont {Qi},\ and\ \citenamefont {Weiner}}]{xue2015mode}%
  \BibitemOpen
  \bibfield  {author} {\bibinfo {author} {\bibfnamefont {X.}~\bibnamefont
  {Xue}}, \bibinfo {author} {\bibfnamefont {Y.}~\bibnamefont {Xuan}}, \bibinfo
  {author} {\bibfnamefont {Y.}~\bibnamefont {Liu}}, \bibinfo {author}
  {\bibfnamefont {P.-H.}\ \bibnamefont {Wang}}, \bibinfo {author}
  {\bibfnamefont {S.}~\bibnamefont {Chen}}, \bibinfo {author} {\bibfnamefont
  {J.}~\bibnamefont {Wang}}, \bibinfo {author} {\bibfnamefont {D.~E.}\
  \bibnamefont {Leaird}}, \bibinfo {author} {\bibfnamefont {M.}~\bibnamefont
  {Qi}}, \ and\ \bibinfo {author} {\bibfnamefont {A.~M.}\ \bibnamefont
  {Weiner}},\ }\href@noop {} {\bibfield  {journal} {\bibinfo  {journal} {Nature
  Photonics}\ }\textbf {\bibinfo {volume} {9}},\ \bibinfo {pages} {594}
  (\bibinfo {year} {2015})}\BibitemShut {NoStop}%
\bibitem [{\citenamefont {Cole}\ \emph {et~al.}(2017)\citenamefont {Cole},
  \citenamefont {Lamb}, \citenamefont {Del'Haye}, \citenamefont {Diddams},\
  and\ \citenamefont {Papp}}]{cole2017soliton}%
  \BibitemOpen
  \bibfield  {author} {\bibinfo {author} {\bibfnamefont {D.~C.}\ \bibnamefont
  {Cole}}, \bibinfo {author} {\bibfnamefont {E.~S.}\ \bibnamefont {Lamb}},
  \bibinfo {author} {\bibfnamefont {P.}~\bibnamefont {Del'Haye}}, \bibinfo
  {author} {\bibfnamefont {S.~A.}\ \bibnamefont {Diddams}}, \ and\ \bibinfo
  {author} {\bibfnamefont {S.~B.}\ \bibnamefont {Papp}},\ }\href@noop {}
  {\bibfield  {journal} {\bibinfo  {journal} {Nature Photonics}\ }\textbf
  {\bibinfo {volume} {11}},\ \bibinfo {pages} {671} (\bibinfo {year}
  {2017})}\BibitemShut {NoStop}%
\bibitem [{\citenamefont {Marin-Palomo}\ \emph {et~al.}(2017)\citenamefont
  {Marin-Palomo}, \citenamefont {Kemal}, \citenamefont {Karpov}, \citenamefont
  {Kordts}, \citenamefont {Pfeifle}, \citenamefont {Pfeiffer}, \citenamefont
  {Trocha}, \citenamefont {Wolf}, \citenamefont {Brasch}, \citenamefont
  {Anderson}, \citenamefont {Rosenberger}, \citenamefont {Vijayan},
  \citenamefont {Freude}, \citenamefont {Kippenberg},\ and\ \citenamefont
  {Koos}}]{marin2017microresonator}%
  \BibitemOpen
  \bibfield  {author} {\bibinfo {author} {\bibfnamefont {P.}~\bibnamefont
  {Marin-Palomo}}, \bibinfo {author} {\bibfnamefont {J.~N.}\ \bibnamefont
  {Kemal}}, \bibinfo {author} {\bibfnamefont {M.}~\bibnamefont {Karpov}},
  \bibinfo {author} {\bibfnamefont {A.}~\bibnamefont {Kordts}}, \bibinfo
  {author} {\bibfnamefont {J.}~\bibnamefont {Pfeifle}}, \bibinfo {author}
  {\bibfnamefont {M.~H.}\ \bibnamefont {Pfeiffer}}, \bibinfo {author}
  {\bibfnamefont {P.}~\bibnamefont {Trocha}}, \bibinfo {author} {\bibfnamefont
  {S.}~\bibnamefont {Wolf}}, \bibinfo {author} {\bibfnamefont {V.}~\bibnamefont
  {Brasch}}, \bibinfo {author} {\bibfnamefont {M.~H.}\ \bibnamefont
  {Anderson}}, \bibinfo {author} {\bibfnamefont {R.}~\bibnamefont
  {Rosenberger}}, \bibinfo {author} {\bibfnamefont {K.}~\bibnamefont
  {Vijayan}}, \bibinfo {author} {\bibfnamefont {W.}~\bibnamefont {Freude}},
  \bibinfo {author} {\bibfnamefont {T.~J.}\ \bibnamefont {Kippenberg}}, \ and\
  \bibinfo {author} {\bibfnamefont {C.}~\bibnamefont {Koos}},\ }\href@noop {}
  {\bibfield  {journal} {\bibinfo  {journal} {Nature}\ }\textbf {\bibinfo
  {volume} {546}},\ \bibinfo {pages} {274} (\bibinfo {year}
  {2017})}\BibitemShut {NoStop}%
\bibitem [{\citenamefont {Spencer}\ \emph {et~al.}(2018)\citenamefont
  {Spencer}, \citenamefont {Drake}, \citenamefont {Briles}, \citenamefont
  {Stone}, \citenamefont {Sinclair}, \citenamefont {Fredrick}, \citenamefont
  {Li}, \citenamefont {Westly}, \citenamefont {Ilic}, \citenamefont
  {Bluestone}, \citenamefont {Volet}, \citenamefont {Komljenovic},
  \citenamefont {Chang}, \citenamefont {Lee}, \citenamefont {Oh}, \citenamefont
  {Suh}, \citenamefont {Yang}, \citenamefont {Pfeiffer}, \citenamefont
  {Kippenberg}, \citenamefont {Norberg}, \citenamefont {Theogarajan},
  \citenamefont {Vahala}, \citenamefont {Newbury}, \citenamefont {Srinivasan},
  \citenamefont {Bowers}, \citenamefont {Diddams},\ and\ \citenamefont
  {Papp}}]{spencer2018integrated}%
  \BibitemOpen
  \bibfield  {author} {\bibinfo {author} {\bibfnamefont {D.~T.}\ \bibnamefont
  {Spencer}}, \bibinfo {author} {\bibfnamefont {T.}~\bibnamefont {Drake}},
  \bibinfo {author} {\bibfnamefont {T.~C.}\ \bibnamefont {Briles}}, \bibinfo
  {author} {\bibfnamefont {J.}~\bibnamefont {Stone}}, \bibinfo {author}
  {\bibfnamefont {L.~C.}\ \bibnamefont {Sinclair}}, \bibinfo {author}
  {\bibfnamefont {C.}~\bibnamefont {Fredrick}}, \bibinfo {author}
  {\bibfnamefont {Q.}~\bibnamefont {Li}}, \bibinfo {author} {\bibfnamefont
  {D.}~\bibnamefont {Westly}}, \bibinfo {author} {\bibfnamefont {B.~R.}\
  \bibnamefont {Ilic}}, \bibinfo {author} {\bibfnamefont {A.}~\bibnamefont
  {Bluestone}}, \bibinfo {author} {\bibfnamefont {N.}~\bibnamefont {Volet}},
  \bibinfo {author} {\bibfnamefont {T.}~\bibnamefont {Komljenovic}}, \bibinfo
  {author} {\bibfnamefont {L.}~\bibnamefont {Chang}}, \bibinfo {author}
  {\bibfnamefont {S.~H.}\ \bibnamefont {Lee}}, \bibinfo {author} {\bibfnamefont
  {D.~Y.}\ \bibnamefont {Oh}}, \bibinfo {author} {\bibfnamefont {M.-G.}\
  \bibnamefont {Suh}}, \bibinfo {author} {\bibfnamefont {K.~Y.}\ \bibnamefont
  {Yang}}, \bibinfo {author} {\bibfnamefont {M.~H.~P.}\ \bibnamefont
  {Pfeiffer}}, \bibinfo {author} {\bibfnamefont {T.~J.}\ \bibnamefont
  {Kippenberg}}, \bibinfo {author} {\bibfnamefont {E.}~\bibnamefont {Norberg}},
  \bibinfo {author} {\bibfnamefont {L.}~\bibnamefont {Theogarajan}}, \bibinfo
  {author} {\bibfnamefont {K.}~\bibnamefont {Vahala}}, \bibinfo {author}
  {\bibfnamefont {N.~R.}\ \bibnamefont {Newbury}}, \bibinfo {author}
  {\bibfnamefont {K.}~\bibnamefont {Srinivasan}}, \bibinfo {author}
  {\bibfnamefont {J.~E.}\ \bibnamefont {Bowers}}, \bibinfo {author}
  {\bibfnamefont {S.~A.}\ \bibnamefont {Diddams}}, \ and\ \bibinfo {author}
  {\bibfnamefont {S.~B.}\ \bibnamefont {Papp}},\ }\href@noop {} {\bibfield
  {journal} {\bibinfo  {journal} {Nature}\ }\textbf {\bibinfo {volume} {557}},\
  \bibinfo {pages} {81} (\bibinfo {year} {2018})}\BibitemShut {NoStop}%
\bibitem [{\citenamefont {Yi}\ \emph {et~al.}(2015)\citenamefont {Yi},
  \citenamefont {Yang}, \citenamefont {Yang}, \citenamefont {Suh},\ and\
  \citenamefont {Vahala}}]{yi2015soliton}%
  \BibitemOpen
  \bibfield  {author} {\bibinfo {author} {\bibfnamefont {X.}~\bibnamefont
  {Yi}}, \bibinfo {author} {\bibfnamefont {Q.-F.}\ \bibnamefont {Yang}},
  \bibinfo {author} {\bibfnamefont {K.~Y.}\ \bibnamefont {Yang}}, \bibinfo
  {author} {\bibfnamefont {M.-G.}\ \bibnamefont {Suh}}, \ and\ \bibinfo
  {author} {\bibfnamefont {K.}~\bibnamefont {Vahala}},\ }\href@noop {}
  {\bibfield  {journal} {\bibinfo  {journal} {Optica}\ }\textbf {\bibinfo
  {volume} {2}},\ \bibinfo {pages} {1078} (\bibinfo {year} {2015})}\BibitemShut
  {NoStop}%
\bibitem [{\citenamefont {Wang}\ \emph {et~al.}(2017)\citenamefont {Wang},
  \citenamefont {Leo}, \citenamefont {Fatome}, \citenamefont {Erkintalo},
  \citenamefont {Murdoch},\ and\ \citenamefont {Coen}}]{wang2017universal}%
  \BibitemOpen
  \bibfield  {author} {\bibinfo {author} {\bibfnamefont {Y.}~\bibnamefont
  {Wang}}, \bibinfo {author} {\bibfnamefont {F.}~\bibnamefont {Leo}}, \bibinfo
  {author} {\bibfnamefont {J.}~\bibnamefont {Fatome}}, \bibinfo {author}
  {\bibfnamefont {M.}~\bibnamefont {Erkintalo}}, \bibinfo {author}
  {\bibfnamefont {S.~G.}\ \bibnamefont {Murdoch}}, \ and\ \bibinfo {author}
  {\bibfnamefont {S.}~\bibnamefont {Coen}},\ }\href@noop {} {\bibfield
  {journal} {\bibinfo  {journal} {Optica}\ }\textbf {\bibinfo {volume} {4}},\
  \bibinfo {pages} {855} (\bibinfo {year} {2017})}\BibitemShut {NoStop}%
\bibitem [{\citenamefont {Matsko}\ \emph {et~al.}(2016)\citenamefont {Matsko},
  \citenamefont {Liang}, \citenamefont {Savchenkov}, \citenamefont {Eliyahu},\
  and\ \citenamefont {Maleki}}]{matsko2016optical}%
  \BibitemOpen
  \bibfield  {author} {\bibinfo {author} {\bibfnamefont {A.~B.}\ \bibnamefont
  {Matsko}}, \bibinfo {author} {\bibfnamefont {W.}~\bibnamefont {Liang}},
  \bibinfo {author} {\bibfnamefont {A.~A.}\ \bibnamefont {Savchenkov}},
  \bibinfo {author} {\bibfnamefont {D.}~\bibnamefont {Eliyahu}}, \ and\
  \bibinfo {author} {\bibfnamefont {L.}~\bibnamefont {Maleki}},\ }\href@noop {}
  {\bibfield  {journal} {\bibinfo  {journal} {Optics letters}\ }\textbf
  {\bibinfo {volume} {41}},\ \bibinfo {pages} {2907} (\bibinfo {year}
  {2016})}\BibitemShut {NoStop}%
\bibitem [{\citenamefont {Brasch}\ \emph {et~al.}(2016)\citenamefont {Brasch},
  \citenamefont {Geiselmann}, \citenamefont {Herr}, \citenamefont {Lihachev},
  \citenamefont {Pfeiffer}, \citenamefont {Gorodetsky},\ and\ \citenamefont
  {Kippenberg}}]{brasch2016photonic}%
  \BibitemOpen
  \bibfield  {author} {\bibinfo {author} {\bibfnamefont {V.}~\bibnamefont
  {Brasch}}, \bibinfo {author} {\bibfnamefont {M.}~\bibnamefont {Geiselmann}},
  \bibinfo {author} {\bibfnamefont {T.}~\bibnamefont {Herr}}, \bibinfo {author}
  {\bibfnamefont {G.}~\bibnamefont {Lihachev}}, \bibinfo {author}
  {\bibfnamefont {M.~H.}\ \bibnamefont {Pfeiffer}}, \bibinfo {author}
  {\bibfnamefont {M.~L.}\ \bibnamefont {Gorodetsky}}, \ and\ \bibinfo {author}
  {\bibfnamefont {T.~J.}\ \bibnamefont {Kippenberg}},\ }\href@noop {}
  {\bibfield  {journal} {\bibinfo  {journal} {Science}\ }\textbf {\bibinfo
  {volume} {351}},\ \bibinfo {pages} {357} (\bibinfo {year}
  {2016})}\BibitemShut {NoStop}%
\bibitem [{\citenamefont {Jaramillo-Villegas}\ \emph
  {et~al.}(2015)\citenamefont {Jaramillo-Villegas}, \citenamefont {Xue},
  \citenamefont {Wang}, \citenamefont {Leaird},\ and\ \citenamefont
  {Weiner}}]{jaramillo2015deterministic}%
  \BibitemOpen
  \bibfield  {author} {\bibinfo {author} {\bibfnamefont {J.~A.}\ \bibnamefont
  {Jaramillo-Villegas}}, \bibinfo {author} {\bibfnamefont {X.}~\bibnamefont
  {Xue}}, \bibinfo {author} {\bibfnamefont {P.-H.}\ \bibnamefont {Wang}},
  \bibinfo {author} {\bibfnamefont {D.~E.}\ \bibnamefont {Leaird}}, \ and\
  \bibinfo {author} {\bibfnamefont {A.~M.}\ \bibnamefont {Weiner}},\
  }\href@noop {} {\bibfield  {journal} {\bibinfo  {journal} {Optics Express}\
  }\textbf {\bibinfo {volume} {23}},\ \bibinfo {pages} {9618} (\bibinfo {year}
  {2015})}\BibitemShut {NoStop}%
\bibitem [{\citenamefont {Briles}\ \emph
  {et~al.}(2018{\natexlab{a}})\citenamefont {Briles}, \citenamefont {Stone},
  \citenamefont {Drake}, \citenamefont {Spencer}, \citenamefont {Fredrick},
  \citenamefont {Li}, \citenamefont {Westly}, \citenamefont {Ilic},
  \citenamefont {Srinivasan}, \citenamefont {Diddams},\ and\ \citenamefont
  {Papp}}]{briles2018interlocking}%
  \BibitemOpen
  \bibfield  {author} {\bibinfo {author} {\bibfnamefont {T.~C.}\ \bibnamefont
  {Briles}}, \bibinfo {author} {\bibfnamefont {J.~R.}\ \bibnamefont {Stone}},
  \bibinfo {author} {\bibfnamefont {T.~E.}\ \bibnamefont {Drake}}, \bibinfo
  {author} {\bibfnamefont {D.~T.}\ \bibnamefont {Spencer}}, \bibinfo {author}
  {\bibfnamefont {C.}~\bibnamefont {Fredrick}}, \bibinfo {author}
  {\bibfnamefont {Q.}~\bibnamefont {Li}}, \bibinfo {author} {\bibfnamefont
  {D.}~\bibnamefont {Westly}}, \bibinfo {author} {\bibfnamefont
  {B.}~\bibnamefont {Ilic}}, \bibinfo {author} {\bibfnamefont {K.}~\bibnamefont
  {Srinivasan}}, \bibinfo {author} {\bibfnamefont {S.~A.}\ \bibnamefont
  {Diddams}}, \ and\ \bibinfo {author} {\bibfnamefont {S.~B.}\ \bibnamefont
  {Papp}},\ }\href@noop {} {\bibfield  {journal} {\bibinfo  {journal} {Optics
  Letters}\ }\textbf {\bibinfo {volume} {43}},\ \bibinfo {pages} {2933}
  (\bibinfo {year} {2018}{\natexlab{a}})}\BibitemShut {NoStop}%
\bibitem [{\citenamefont {Stern}\ \emph {et~al.}(2018)\citenamefont {Stern},
  \citenamefont {Ji}, \citenamefont {Okawachi}, \citenamefont {Gaeta},\ and\
  \citenamefont {Lipson}}]{stern2018battery}%
  \BibitemOpen
  \bibfield  {author} {\bibinfo {author} {\bibfnamefont {B.}~\bibnamefont
  {Stern}}, \bibinfo {author} {\bibfnamefont {X.}~\bibnamefont {Ji}}, \bibinfo
  {author} {\bibfnamefont {Y.}~\bibnamefont {Okawachi}}, \bibinfo {author}
  {\bibfnamefont {A.~L.}\ \bibnamefont {Gaeta}}, \ and\ \bibinfo {author}
  {\bibfnamefont {M.}~\bibnamefont {Lipson}},\ }\href@noop {} {\bibfield
  {journal} {\bibinfo  {journal} {Nature}\ ,\ \bibinfo {pages} {1}} (\bibinfo
  {year} {2018})}\BibitemShut {NoStop}%
\bibitem [{\citenamefont {Papp}\ \emph {et~al.}(2014)\citenamefont {Papp},
  \citenamefont {Beha}, \citenamefont {Del'Haye}, \citenamefont {Quinlan},
  \citenamefont {Lee}, \citenamefont {Vahala},\ and\ \citenamefont
  {Diddams}}]{papp2014microresonator}%
  \BibitemOpen
  \bibfield  {author} {\bibinfo {author} {\bibfnamefont {S.~B.}\ \bibnamefont
  {Papp}}, \bibinfo {author} {\bibfnamefont {K.}~\bibnamefont {Beha}}, \bibinfo
  {author} {\bibfnamefont {P.}~\bibnamefont {Del'Haye}}, \bibinfo {author}
  {\bibfnamefont {F.}~\bibnamefont {Quinlan}}, \bibinfo {author} {\bibfnamefont
  {H.}~\bibnamefont {Lee}}, \bibinfo {author} {\bibfnamefont {K.~J.}\
  \bibnamefont {Vahala}}, \ and\ \bibinfo {author} {\bibfnamefont {S.~A.}\
  \bibnamefont {Diddams}},\ }\href@noop {} {\bibfield  {journal} {\bibinfo
  {journal} {Optica}\ }\textbf {\bibinfo {volume} {1}},\ \bibinfo {pages} {10}
  (\bibinfo {year} {2014})}\BibitemShut {NoStop}%
\bibitem [{\citenamefont {Karpov}\ \emph {et~al.}(2018)\citenamefont {Karpov},
  \citenamefont {Pfeiffer}, \citenamefont {Liu}, \citenamefont {Lukashchuk},\
  and\ \citenamefont {Kippenberg}}]{Karpov2018}%
  \BibitemOpen
  \bibfield  {author} {\bibinfo {author} {\bibfnamefont {M.}~\bibnamefont
  {Karpov}}, \bibinfo {author} {\bibfnamefont {M.~H.~P.}\ \bibnamefont
  {Pfeiffer}}, \bibinfo {author} {\bibfnamefont {J.}~\bibnamefont {Liu}},
  \bibinfo {author} {\bibfnamefont {A.}~\bibnamefont {Lukashchuk}}, \ and\
  \bibinfo {author} {\bibfnamefont {T.~J.}\ \bibnamefont {Kippenberg}},\ }\href
  {\doibase 10.1038/s41467-018-03471-x} {\bibfield  {journal} {\bibinfo
  {journal} {Nature Communications}\ }\textbf {\bibinfo {volume} {9}} (\bibinfo
  {year} {2018}),\ 10.1038/s41467-018-03471-x}\BibitemShut {NoStop}%
\bibitem [{\citenamefont {Agha}\ \emph {et~al.}(2009)\citenamefont {Agha},
  \citenamefont {Okawachi},\ and\ \citenamefont {Gaeta}}]{agha2009theoretical}%
  \BibitemOpen
  \bibfield  {author} {\bibinfo {author} {\bibfnamefont {I.~H.}\ \bibnamefont
  {Agha}}, \bibinfo {author} {\bibfnamefont {Y.}~\bibnamefont {Okawachi}}, \
  and\ \bibinfo {author} {\bibfnamefont {A.~L.}\ \bibnamefont {Gaeta}},\
  }\href@noop {} {\bibfield  {journal} {\bibinfo  {journal} {Optics Express}\
  }\textbf {\bibinfo {volume} {17}},\ \bibinfo {pages} {16209} (\bibinfo {year}
  {2009})}\BibitemShut {NoStop}%
\bibitem [{\citenamefont {Okawachi}\ \emph {et~al.}(2014)\citenamefont
  {Okawachi}, \citenamefont {Lamont}, \citenamefont {Luke}, \citenamefont
  {Carvalho}, \citenamefont {Yu}, \citenamefont {Lipson},\ and\ \citenamefont
  {Gaeta}}]{okawachi2014bandwidth}%
  \BibitemOpen
  \bibfield  {author} {\bibinfo {author} {\bibfnamefont {Y.}~\bibnamefont
  {Okawachi}}, \bibinfo {author} {\bibfnamefont {M.~R.}\ \bibnamefont
  {Lamont}}, \bibinfo {author} {\bibfnamefont {K.}~\bibnamefont {Luke}},
  \bibinfo {author} {\bibfnamefont {D.~O.}\ \bibnamefont {Carvalho}}, \bibinfo
  {author} {\bibfnamefont {M.}~\bibnamefont {Yu}}, \bibinfo {author}
  {\bibfnamefont {M.}~\bibnamefont {Lipson}}, \ and\ \bibinfo {author}
  {\bibfnamefont {A.~L.}\ \bibnamefont {Gaeta}},\ }\href@noop {} {\bibfield
  {journal} {\bibinfo  {journal} {Optics Letters}\ }\textbf {\bibinfo {volume}
  {39}},\ \bibinfo {pages} {3535} (\bibinfo {year} {2014})}\BibitemShut
  {NoStop}%
\bibitem [{\citenamefont {Diddams}\ \emph {et~al.}(2004)\citenamefont
  {Diddams}, \citenamefont {Bergquist}, \citenamefont {Jefferts},\ and\
  \citenamefont {Oates}}]{diddams2004standards}%
  \BibitemOpen
  \bibfield  {author} {\bibinfo {author} {\bibfnamefont {S.~A.}\ \bibnamefont
  {Diddams}}, \bibinfo {author} {\bibfnamefont {J.~C.}\ \bibnamefont
  {Bergquist}}, \bibinfo {author} {\bibfnamefont {S.~R.}\ \bibnamefont
  {Jefferts}}, \ and\ \bibinfo {author} {\bibfnamefont {C.~W.}\ \bibnamefont
  {Oates}},\ }\href@noop {} {\bibfield  {journal} {\bibinfo  {journal}
  {Science}\ }\textbf {\bibinfo {volume} {306}},\ \bibinfo {pages} {1318}
  (\bibinfo {year} {2004})}\BibitemShut {NoStop}%
\bibitem [{\citenamefont {Ludlow}\ \emph {et~al.}(2015)\citenamefont {Ludlow},
  \citenamefont {Boyd}, \citenamefont {Ye}, \citenamefont {Peik},\ and\
  \citenamefont {Schmidt}}]{ludlow2015optical}%
  \BibitemOpen
  \bibfield  {author} {\bibinfo {author} {\bibfnamefont {A.~D.}\ \bibnamefont
  {Ludlow}}, \bibinfo {author} {\bibfnamefont {M.~M.}\ \bibnamefont {Boyd}},
  \bibinfo {author} {\bibfnamefont {J.}~\bibnamefont {Ye}}, \bibinfo {author}
  {\bibfnamefont {E.}~\bibnamefont {Peik}}, \ and\ \bibinfo {author}
  {\bibfnamefont {P.~O.}\ \bibnamefont {Schmidt}},\ }\href@noop {} {\bibfield
  {journal} {\bibinfo  {journal} {Reviews of Modern Physics}\ }\textbf
  {\bibinfo {volume} {87}},\ \bibinfo {pages} {637} (\bibinfo {year}
  {2015})}\BibitemShut {NoStop}%
\bibitem [{\citenamefont {Targat}\ \emph {et~al.}(2013)\citenamefont {Targat},
  \citenamefont {Lorini}, \citenamefont {Coq}, \citenamefont {Zawada},
  \citenamefont {Guéna}, \citenamefont {Abgrall}, \citenamefont {Gurov},
  \citenamefont {Rosenbusch}, \citenamefont {Rovera}, \citenamefont {Nagórny},
  \citenamefont {Gartman}, \citenamefont {Westergaard}, \citenamefont {Tobar},
  \citenamefont {Lours}, \citenamefont {Santarelli}, \citenamefont {Clairon},
  \citenamefont {Bize}, \citenamefont {Laurent}, \citenamefont {Lemonde},\ and\
  \citenamefont {Lodewyck}}]{Targat2013}%
  \BibitemOpen
  \bibfield  {author} {\bibinfo {author} {\bibfnamefont {R.~L.}\ \bibnamefont
  {Targat}}, \bibinfo {author} {\bibfnamefont {L.}~\bibnamefont {Lorini}},
  \bibinfo {author} {\bibfnamefont {Y.~L.}\ \bibnamefont {Coq}}, \bibinfo
  {author} {\bibfnamefont {M.}~\bibnamefont {Zawada}}, \bibinfo {author}
  {\bibfnamefont {J.}~\bibnamefont {Guéna}}, \bibinfo {author} {\bibfnamefont
  {M.}~\bibnamefont {Abgrall}}, \bibinfo {author} {\bibfnamefont
  {M.}~\bibnamefont {Gurov}}, \bibinfo {author} {\bibfnamefont
  {P.}~\bibnamefont {Rosenbusch}}, \bibinfo {author} {\bibfnamefont {D.~G.}\
  \bibnamefont {Rovera}}, \bibinfo {author} {\bibfnamefont {B.}~\bibnamefont
  {Nagórny}}, \bibinfo {author} {\bibfnamefont {R.}~\bibnamefont {Gartman}},
  \bibinfo {author} {\bibfnamefont {P.~G.}\ \bibnamefont {Westergaard}},
  \bibinfo {author} {\bibfnamefont {M.~E.}\ \bibnamefont {Tobar}}, \bibinfo
  {author} {\bibfnamefont {M.}~\bibnamefont {Lours}}, \bibinfo {author}
  {\bibfnamefont {G.}~\bibnamefont {Santarelli}}, \bibinfo {author}
  {\bibfnamefont {A.}~\bibnamefont {Clairon}}, \bibinfo {author} {\bibfnamefont
  {S.}~\bibnamefont {Bize}}, \bibinfo {author} {\bibfnamefont {P.}~\bibnamefont
  {Laurent}}, \bibinfo {author} {\bibfnamefont {P.}~\bibnamefont {Lemonde}}, \
  and\ \bibinfo {author} {\bibfnamefont {J.}~\bibnamefont {Lodewyck}},\ }\href
  {\doibase 10.1038/ncomms3109} {\bibfield  {journal} {\bibinfo  {journal}
  {Nature Communications}\ }\textbf {\bibinfo {volume} {4}} (\bibinfo {year}
  {2013}),\ 10.1038/ncomms3109}\BibitemShut {NoStop}%
\bibitem [{\citenamefont {Skryabin}\ and\ \citenamefont
  {Kartashov}(2017)}]{skryabin2017self}%
  \BibitemOpen
  \bibfield  {author} {\bibinfo {author} {\bibfnamefont {D.}~\bibnamefont
  {Skryabin}}\ and\ \bibinfo {author} {\bibfnamefont {Y.}~\bibnamefont
  {Kartashov}},\ }\href@noop {} {\bibfield  {journal} {\bibinfo  {journal}
  {Optics Express}\ }\textbf {\bibinfo {volume} {25}},\ \bibinfo {pages}
  {27442} (\bibinfo {year} {2017})}\BibitemShut {NoStop}%
\bibitem [{\citenamefont {Hilico}\ \emph {et~al.}(1998)\citenamefont {Hilico},
  \citenamefont {Felder}, \citenamefont {Touahri}, \citenamefont {Acef},
  \citenamefont {Clairon},\ and\ \citenamefont
  {Biraben}}]{hilico1998metrological}%
  \BibitemOpen
  \bibfield  {author} {\bibinfo {author} {\bibfnamefont {L.}~\bibnamefont
  {Hilico}}, \bibinfo {author} {\bibfnamefont {R.}~\bibnamefont {Felder}},
  \bibinfo {author} {\bibfnamefont {D.}~\bibnamefont {Touahri}}, \bibinfo
  {author} {\bibfnamefont {O.}~\bibnamefont {Acef}}, \bibinfo {author}
  {\bibfnamefont {A.}~\bibnamefont {Clairon}}, \ and\ \bibinfo {author}
  {\bibfnamefont {F.}~\bibnamefont {Biraben}},\ }\href@noop {} {\bibfield
  {journal} {\bibinfo  {journal} {The European Physical Journal-Applied
  Physics}\ }\textbf {\bibinfo {volume} {4}},\ \bibinfo {pages} {219} (\bibinfo
  {year} {1998})}\BibitemShut {NoStop}%
\bibitem [{\citenamefont {Zameroski}\ \emph {et~al.}(2014)\citenamefont
  {Zameroski}, \citenamefont {Hager}, \citenamefont {Erickson},\ and\
  \citenamefont {Burke}}]{zameroski2014pressure}%
  \BibitemOpen
  \bibfield  {author} {\bibinfo {author} {\bibfnamefont {N.~D.}\ \bibnamefont
  {Zameroski}}, \bibinfo {author} {\bibfnamefont {G.~D.}\ \bibnamefont
  {Hager}}, \bibinfo {author} {\bibfnamefont {C.~J.}\ \bibnamefont {Erickson}},
  \ and\ \bibinfo {author} {\bibfnamefont {J.~H.}\ \bibnamefont {Burke}},\
  }\href {http://stacks.iop.org/0953-4075/47/i=22/a=225205} {\bibfield
  {journal} {\bibinfo  {journal} {Journal of Physics B: Atomic, Molecular and
  Optical Physics}\ }\textbf {\bibinfo {volume} {47}},\ \bibinfo {pages}
  {225205} (\bibinfo {year} {2014})}\BibitemShut {NoStop}%
\bibitem [{\citenamefont {Bernard}\ \emph {et~al.}(2000)\citenamefont
  {Bernard}, \citenamefont {Madej}, \citenamefont {Siemsen}, \citenamefont
  {Marmet}, \citenamefont {Latrasse}, \citenamefont {Touahri}, \citenamefont
  {Poulin}, \citenamefont {Allard},\ and\ \citenamefont
  {T{\^e}tu}}]{bernard2000absolute}%
  \BibitemOpen
  \bibfield  {author} {\bibinfo {author} {\bibfnamefont {J.}~\bibnamefont
  {Bernard}}, \bibinfo {author} {\bibfnamefont {A.}~\bibnamefont {Madej}},
  \bibinfo {author} {\bibfnamefont {K.}~\bibnamefont {Siemsen}}, \bibinfo
  {author} {\bibfnamefont {L.}~\bibnamefont {Marmet}}, \bibinfo {author}
  {\bibfnamefont {C.}~\bibnamefont {Latrasse}}, \bibinfo {author}
  {\bibfnamefont {D.}~\bibnamefont {Touahri}}, \bibinfo {author} {\bibfnamefont
  {M.}~\bibnamefont {Poulin}}, \bibinfo {author} {\bibfnamefont
  {M.}~\bibnamefont {Allard}}, \ and\ \bibinfo {author} {\bibfnamefont
  {M.}~\bibnamefont {T{\^e}tu}},\ }\href@noop {} {\bibfield  {journal}
  {\bibinfo  {journal} {Optics Communications}\ }\textbf {\bibinfo {volume}
  {173}},\ \bibinfo {pages} {357} (\bibinfo {year} {2000})}\BibitemShut
  {NoStop}%
\bibitem [{\citenamefont {Hosseini}\ \emph {et~al.}(2010)\citenamefont
  {Hosseini}, \citenamefont {Yegnanarayanan}, \citenamefont {Atabaki},
  \citenamefont {Soltani},\ and\ \citenamefont
  {Adibi}}]{hosseini2010systematic}%
  \BibitemOpen
  \bibfield  {author} {\bibinfo {author} {\bibfnamefont {E.~S.}\ \bibnamefont
  {Hosseini}}, \bibinfo {author} {\bibfnamefont {S.}~\bibnamefont
  {Yegnanarayanan}}, \bibinfo {author} {\bibfnamefont {A.~H.}\ \bibnamefont
  {Atabaki}}, \bibinfo {author} {\bibfnamefont {M.}~\bibnamefont {Soltani}}, \
  and\ \bibinfo {author} {\bibfnamefont {A.}~\bibnamefont {Adibi}},\
  }\href@noop {} {\bibfield  {journal} {\bibinfo  {journal} {Optics Express}\
  }\textbf {\bibinfo {volume} {18}},\ \bibinfo {pages} {2127} (\bibinfo {year}
  {2010})}\BibitemShut {NoStop}%
\bibitem [{\citenamefont {Li}\ \emph {et~al.}(2015)\citenamefont {Li},
  \citenamefont {Briles}, \citenamefont {Westly}, \citenamefont {Stone},
  \citenamefont {Ilic}, \citenamefont {Diddams}, \citenamefont {Papp},\ and\
  \citenamefont {Srinivasan}}]{li2015octave}%
  \BibitemOpen
  \bibfield  {author} {\bibinfo {author} {\bibfnamefont {Q.}~\bibnamefont
  {Li}}, \bibinfo {author} {\bibfnamefont {T.~C.}\ \bibnamefont {Briles}},
  \bibinfo {author} {\bibfnamefont {D.}~\bibnamefont {Westly}}, \bibinfo
  {author} {\bibfnamefont {J.}~\bibnamefont {Stone}}, \bibinfo {author}
  {\bibfnamefont {R.}~\bibnamefont {Ilic}}, \bibinfo {author} {\bibfnamefont
  {S.}~\bibnamefont {Diddams}}, \bibinfo {author} {\bibfnamefont
  {S.}~\bibnamefont {Papp}}, \ and\ \bibinfo {author} {\bibfnamefont
  {K.}~\bibnamefont {Srinivasan}},\ }in\ \href@noop {} {\emph {\bibinfo
  {booktitle} {Frontiers in Optics}}}\ (\bibinfo {organization} {Optical
  Society of America},\ \bibinfo {year} {2015})\ pp.\ \bibinfo {pages}
  {FW6C--5}\BibitemShut {NoStop}%
\bibitem [{\citenamefont {Briles}\ \emph
  {et~al.}(2018{\natexlab{b}})\citenamefont {Briles}, \citenamefont {Stone},
  \citenamefont {Drake}, \citenamefont {Spencer}, \citenamefont {Diddams},\
  and\ \citenamefont {Papp}}]{briles2018thermal}%
  \BibitemOpen
  \bibfield  {author} {\bibinfo {author} {\bibfnamefont {T.~C.}\ \bibnamefont
  {Briles}}, \bibinfo {author} {\bibfnamefont {J.~R.}\ \bibnamefont {Stone}},
  \bibinfo {author} {\bibfnamefont {T.~E.}\ \bibnamefont {Drake}}, \bibinfo
  {author} {\bibfnamefont {D.~T.}\ \bibnamefont {Spencer}}, \bibinfo {author}
  {\bibfnamefont {S.~A.}\ \bibnamefont {Diddams}}, \ and\ \bibinfo {author}
  {\bibfnamefont {S.~B.}\ \bibnamefont {Papp}},\ }\href@noop {} {\bibfield
  {journal} {\bibinfo  {journal} {In preparation}\ } (\bibinfo {year}
  {2018}{\natexlab{b}})}\BibitemShut {NoStop}%
\bibitem [{\citenamefont {Stone}\ \emph {et~al.}(2018)\citenamefont {Stone},
  \citenamefont {Briles}, \citenamefont {Drake}, \citenamefont {Spencer},
  \citenamefont {Carlson}, \citenamefont {Diddams},\ and\ \citenamefont
  {Papp}}]{stone2018thermal}%
  \BibitemOpen
  \bibfield  {author} {\bibinfo {author} {\bibfnamefont {J.~R.}\ \bibnamefont
  {Stone}}, \bibinfo {author} {\bibfnamefont {T.~C.}\ \bibnamefont {Briles}},
  \bibinfo {author} {\bibfnamefont {T.~E.}\ \bibnamefont {Drake}}, \bibinfo
  {author} {\bibfnamefont {D.~T.}\ \bibnamefont {Spencer}}, \bibinfo {author}
  {\bibfnamefont {D.~R.}\ \bibnamefont {Carlson}}, \bibinfo {author}
  {\bibfnamefont {S.~A.}\ \bibnamefont {Diddams}}, \ and\ \bibinfo {author}
  {\bibfnamefont {S.~B.}\ \bibnamefont {Papp}},\ }\href@noop {} {\bibfield
  {journal} {\bibinfo  {journal} {Physical Review Letters}\ }\textbf {\bibinfo
  {volume} {121}},\ \bibinfo {pages} {063902} (\bibinfo {year}
  {2018})}\BibitemShut {NoStop}%
\bibitem [{\citenamefont {Li}\ \emph {et~al.}(2017)\citenamefont {Li},
  \citenamefont {Briles}, \citenamefont {Westly}, \citenamefont {Drake},
  \citenamefont {Stone}, \citenamefont {Ilic}, \citenamefont {Diddams},
  \citenamefont {Papp},\ and\ \citenamefont {Srinivasan}}]{li2017stably}%
  \BibitemOpen
  \bibfield  {author} {\bibinfo {author} {\bibfnamefont {Q.}~\bibnamefont
  {Li}}, \bibinfo {author} {\bibfnamefont {T.~C.}\ \bibnamefont {Briles}},
  \bibinfo {author} {\bibfnamefont {D.~A.}\ \bibnamefont {Westly}}, \bibinfo
  {author} {\bibfnamefont {T.~E.}\ \bibnamefont {Drake}}, \bibinfo {author}
  {\bibfnamefont {J.~R.}\ \bibnamefont {Stone}}, \bibinfo {author}
  {\bibfnamefont {B.~R.}\ \bibnamefont {Ilic}}, \bibinfo {author}
  {\bibfnamefont {S.~A.}\ \bibnamefont {Diddams}}, \bibinfo {author}
  {\bibfnamefont {S.~B.}\ \bibnamefont {Papp}}, \ and\ \bibinfo {author}
  {\bibfnamefont {K.}~\bibnamefont {Srinivasan}},\ }\href@noop {} {\bibfield
  {journal} {\bibinfo  {journal} {Optica}\ }\textbf {\bibinfo {volume} {4}},\
  \bibinfo {pages} {193} (\bibinfo {year} {2017})}\BibitemShut {NoStop}%
\bibitem [{\citenamefont {Luke}\ \emph {et~al.}(2015)\citenamefont {Luke},
  \citenamefont {Okawachi}, \citenamefont {Lamont}, \citenamefont {Gaeta},\
  and\ \citenamefont {Lipson}}]{Luke2015}%
  \BibitemOpen
  \bibfield  {author} {\bibinfo {author} {\bibfnamefont {K.}~\bibnamefont
  {Luke}}, \bibinfo {author} {\bibfnamefont {Y.}~\bibnamefont {Okawachi}},
  \bibinfo {author} {\bibfnamefont {M.~R.~E.}\ \bibnamefont {Lamont}}, \bibinfo
  {author} {\bibfnamefont {A.~L.}\ \bibnamefont {Gaeta}}, \ and\ \bibinfo
  {author} {\bibfnamefont {M.}~\bibnamefont {Lipson}},\ }\href {\doibase
  10.1364/OL.40.004823} {\bibfield  {journal} {\bibinfo  {journal} {Optics
  Letters}\ }\textbf {\bibinfo {volume} {40}} (\bibinfo {year} {2015}),\
  10.1364/OL.40.004823}\BibitemShut {NoStop}%
\bibitem [{\citenamefont {Malitson}(1965)}]{Malitson1965}%
  \BibitemOpen
  \bibfield  {author} {\bibinfo {author} {\bibfnamefont {I.~H.}\ \bibnamefont
  {Malitson}},\ }\href {\doibase 10.1364/JOSA.55.001205} {\bibfield  {journal}
  {\bibinfo  {journal} {Journal of the Optical Society of America}\ }\textbf
  {\bibinfo {volume} {55}},\ \bibinfo {pages} {1205} (\bibinfo {year}
  {1965})}\BibitemShut {NoStop}%
\bibitem [{\citenamefont {Del’Haye}\ \emph {et~al.}(2012)\citenamefont
  {Del’Haye}, \citenamefont {Papp},\ and\ \citenamefont
  {Diddams}}]{del2012hybrid}%
  \BibitemOpen
  \bibfield  {author} {\bibinfo {author} {\bibfnamefont {P.}~\bibnamefont
  {Del’Haye}}, \bibinfo {author} {\bibfnamefont {S.~B.}\ \bibnamefont
  {Papp}}, \ and\ \bibinfo {author} {\bibfnamefont {S.~A.}\ \bibnamefont
  {Diddams}},\ }\href@noop {} {\bibfield  {journal} {\bibinfo  {journal}
  {Physical Review Letters}\ }\textbf {\bibinfo {volume} {109}},\ \bibinfo
  {pages} {263901} (\bibinfo {year} {2012})}\BibitemShut {NoStop}%
\bibitem [{\citenamefont {Heinecke}\ \emph {et~al.}(2009)\citenamefont
  {Heinecke}, \citenamefont {Bartels}, \citenamefont {Fortier}, \citenamefont
  {Braje}, \citenamefont {Hollberg},\ and\ \citenamefont
  {Diddams}}]{Heinecke2009}%
  \BibitemOpen
  \bibfield  {author} {\bibinfo {author} {\bibfnamefont {D.~C.}\ \bibnamefont
  {Heinecke}}, \bibinfo {author} {\bibfnamefont {A.}~\bibnamefont {Bartels}},
  \bibinfo {author} {\bibfnamefont {T.~M.}\ \bibnamefont {Fortier}}, \bibinfo
  {author} {\bibfnamefont {D.~A.}\ \bibnamefont {Braje}}, \bibinfo {author}
  {\bibfnamefont {L.}~\bibnamefont {Hollberg}}, \ and\ \bibinfo {author}
  {\bibfnamefont {S.~A.}\ \bibnamefont {Diddams}},\ }\href {\doibase
  10.1103/PhysRevA.80.053806} {\bibfield  {journal} {\bibinfo  {journal}
  {Physical Review A}\ }\textbf {\bibinfo {volume} {80}},\ \bibinfo {pages}
  {053806} (\bibinfo {year} {2009})}\BibitemShut {NoStop}%
\bibitem [{\citenamefont {Ikeda}\ \emph {et~al.}(2008)\citenamefont {Ikeda},
  \citenamefont {Saperstein}, \citenamefont {Alic},\ and\ \citenamefont
  {Fainman}}]{ikeda2008thermal}%
  \BibitemOpen
  \bibfield  {author} {\bibinfo {author} {\bibfnamefont {K.}~\bibnamefont
  {Ikeda}}, \bibinfo {author} {\bibfnamefont {R.~E.}\ \bibnamefont
  {Saperstein}}, \bibinfo {author} {\bibfnamefont {N.}~\bibnamefont {Alic}}, \
  and\ \bibinfo {author} {\bibfnamefont {Y.}~\bibnamefont {Fainman}},\
  }\href@noop {} {\bibfield  {journal} {\bibinfo  {journal} {Optics Express}\
  }\textbf {\bibinfo {volume} {16}},\ \bibinfo {pages} {12987} (\bibinfo {year}
  {2008})}\BibitemShut {NoStop}%
\bibitem [{\citenamefont {Xue}\ \emph {et~al.}(2016)\citenamefont {Xue},
  \citenamefont {Xuan}, \citenamefont {Wang}, \citenamefont {Wang},
  \citenamefont {Liu}, \citenamefont {Niu}, \citenamefont {Leaird},
  \citenamefont {Qi},\ and\ \citenamefont {Weiner}}]{xue2016thermal}%
  \BibitemOpen
  \bibfield  {author} {\bibinfo {author} {\bibfnamefont {X.}~\bibnamefont
  {Xue}}, \bibinfo {author} {\bibfnamefont {Y.}~\bibnamefont {Xuan}}, \bibinfo
  {author} {\bibfnamefont {C.}~\bibnamefont {Wang}}, \bibinfo {author}
  {\bibfnamefont {P.-H.}\ \bibnamefont {Wang}}, \bibinfo {author}
  {\bibfnamefont {Y.}~\bibnamefont {Liu}}, \bibinfo {author} {\bibfnamefont
  {B.}~\bibnamefont {Niu}}, \bibinfo {author} {\bibfnamefont {D.~E.}\
  \bibnamefont {Leaird}}, \bibinfo {author} {\bibfnamefont {M.}~\bibnamefont
  {Qi}}, \ and\ \bibinfo {author} {\bibfnamefont {A.~M.}\ \bibnamefont
  {Weiner}},\ }\href@noop {} {\bibfield  {journal} {\bibinfo  {journal} {Optics
  Express}\ }\textbf {\bibinfo {volume} {24}},\ \bibinfo {pages} {687}
  (\bibinfo {year} {2016})}\BibitemShut {NoStop}%
\bibitem [{\citenamefont {Joshi}\ \emph {et~al.}(2016)\citenamefont {Joshi},
  \citenamefont {Jang}, \citenamefont {Luke}, \citenamefont {Ji}, \citenamefont
  {Miller}, \citenamefont {Klenner}, \citenamefont {Okawachi}, \citenamefont
  {Lipson},\ and\ \citenamefont {Gaeta}}]{joshi2016thermally}%
  \BibitemOpen
  \bibfield  {author} {\bibinfo {author} {\bibfnamefont {C.}~\bibnamefont
  {Joshi}}, \bibinfo {author} {\bibfnamefont {J.~K.}\ \bibnamefont {Jang}},
  \bibinfo {author} {\bibfnamefont {K.}~\bibnamefont {Luke}}, \bibinfo {author}
  {\bibfnamefont {X.}~\bibnamefont {Ji}}, \bibinfo {author} {\bibfnamefont
  {S.~A.}\ \bibnamefont {Miller}}, \bibinfo {author} {\bibfnamefont
  {A.}~\bibnamefont {Klenner}}, \bibinfo {author} {\bibfnamefont
  {Y.}~\bibnamefont {Okawachi}}, \bibinfo {author} {\bibfnamefont
  {M.}~\bibnamefont {Lipson}}, \ and\ \bibinfo {author} {\bibfnamefont {A.~L.}\
  \bibnamefont {Gaeta}},\ }\href@noop {} {\bibfield  {journal} {\bibinfo
  {journal} {Optics Letters}\ }\textbf {\bibinfo {volume} {41}},\ \bibinfo
  {pages} {2565} (\bibinfo {year} {2016})}\BibitemShut {NoStop}%
\end{thebibliography}%

\end{document}